\documentclass[a4,12pt]{article}

\usepackage{graphicx}
\usepackage{amsmath,amssymb}
\usepackage{rotating}
\usepackage[authoryear]{natbib}

\newcommand\argmax{\operatornamewithlimits{argmax}}
\newcommand\maximize{\mathop{\rm maximize}}
\begin{document}

\title{Classifying real-world data with the $DD\alpha$-procedure}

\author{Pavlo Mozharovskyi$^\ast$\and
        Karl Mosler$^\ast$\and
        Tatjana Lange$^{\ast\ast}$\and
        \phantom{aaaaaaaaaaaaaaaaaaaaaaaaaaaaaaaaaaaaaaaaaaaaaaaaaaaaaaaaaaaaaaaaa}\and
        {\small $^\ast$Universit\"at zu K\"oln, Albertus-Magnus-Platz, 50923 K\"oln, Germany} \and
        {\small $^{\ast\ast}$Hochschule Merseburg, Geusaer Stra\ss e, 06217 Merseburg, Germany}}

\date{August 24, 2014}

\maketitle

\begin{abstract}
The $DD\alpha$-classifier, a nonparametric fast and very robust procedure, is described and applied to fifty classification problems regarding a broad spectrum of real-world data.
The procedure first transforms the data from their original property space into a depth space, which is a low-dimensional unit cube, and then separates them by a projective invariant procedure, called $\alpha$-procedure. To each data point the transformation assigns its depth values with respect to the given classes.
Several alternative depth notions (spatial depth, Mahalanobis depth, projection depth, and Tukey depth, the latter two being approximated by univariate projections) are used in the procedure, and compared regarding their average error rates.
With the Tukey depth, which fits the distributions' shape best and is most robust, `outsiders', that is data points having zero depth in all classes, appear. They need an additional treatment for classification.
Evidence is also given about
the dimension of the extended feature space needed for linear separation.
The $DD\alpha$-procedure is available as an R-package.
\end{abstract}


\section{Introduction}\label{sec:intro}

Many statistical procedures have been developed to classify data into two or more given classes.
Generally, if the data arise from a known class of distributions, properties of the classifiers are established through either theoretical considerations or  simulation studies. By this, alternative classifiers are compared and procedures identified that are optimal under properly chosen assumptions.
However, real-world data do often not fit into standard parametric distribution models.
Classifying them requires nonparametric procedures, while, due to the lack of established general properties, selecting a good classifier has to be mainly based on empirical evidence.
Usually such evidence is sought from simulation studies mimicking certain features of the data that arise in practical applications. At best, in a given field of application so called `stylized facts' are identified and translated into a simulation setting.
But often such `stylized facts' do not exist. Then, we cannot learn much from simulation studies how a statistical procedure really works in practice. The adequacy and fitness of the procedure can only be demonstrated when it is applied to real-world data, and its \emph{general} fitness can be only established by successful application to a \emph{large variety} of such data.

In the sequel this is done for a newly developed nonparametric classifier,
the $DD\alpha$-procedure \citep{LangeMM12a}. It is applied to fifty binary classification problems regarding real-world benchmark data.

The $DD\alpha$-procedure first transforms the data from their original property space into a \emph{depth space}, which is a low-dimensional unit cube, and then separates them by a projective invariant procedure, called \emph{$\alpha$-procedure}. To each data point the transformation assigns its depth values with respect to the $q$ given  classes.
The depth coordinates of the data reflect their degree of centrality w.r.t.\ each of the classes.
This central ordering is carried out using a properly chosen depth function.
The subsequent separation in the depth space accounts only for differences in the depth values: If $q=2$ a binary separator is determined by the $\alpha$-procedure. The $\alpha$-procedure stepwise selects pairs of \emph{extended depth properties} (that is, depth coordinates and powers and products of them) and separates them by a linear rule. The separator is a hyperplane in the extended depth space, which corresponds to a polynomial line in the basic depth plane, both containing the origin as an element.
With $q>2$ classes, $q \choose 2$ such $\alpha$-separations can be performed and a majority rule applied; alternatively $q$ one-against-all separators can be used.
We restrict the present study to the case $q=2$, see~\cite{LangeMM12a} for $q>2$.

In \cite{LangeMM12a} the zonoid depth \citep{KoshevoyM97,Mosler02} is applied, which is efficiently computed also in higher dimensions. Here
we employ four alternative depths: the Mahalanobis depth, the spatial depth, the projection depth and the Tukey depth. The first three depths are positive everywhere, while the Tukey depth (like the zonoid depth) vanishes outside the convex hull of the data. However, the Tukey depth reflects the shape of the data much better than the previous three do and it is more robust against outliers than these (and than the zonoid depth as well).
For computational reasons we use the \emph{random Tukey depth} \citep{CuestaANR08}, which approximates the Tukey (= location) depth by minimizing univariate Tukey depths over a finite number of directions.

When using the random Tukey depth (or another depth that vanishes outside the convex hull of the data) a first practical question is how outsiders, that is data points having zero depth in all classes, should be treated. These points are, by construction, represented by the origin of the depth space and, hence, arbitrarily assigned.
With real data, often a large portion of the data turn out to be such outsiders.
As our task is to classify all points, we need either a depth that does not produce outsiders, or a supplementary treatment of outsiders. By definition, the $DD\alpha$-procedure includes a treatment of outsiders if necessary.
For the $DD\alpha$-classifier with the random Tukey depth, several possible treatments are introduced in the sequel. The paper considers the respective variants of the  $DD\alpha$-classifier and compares them with the $DD\alpha$-classifiers based on Mahalanobis, spatial and projection depths.
Recall that the latter depths, as they are positive on the entire $\mathbb{R}^d$, do not yield outsiders.

A second question is how many directions should be chosen to approximate the Tukey depth and how they should be generated; it is addressed in \cite{LangeMM12b} by means of a simulation study.
(For the random projection depth this question is less important as it has no outsiders.)
Broad numerical experience is provided about the relative usefulness of the classifiers.
Further we investigate how many features in the extended depth space are needed on an average to satisfactorily separate the data.
Finally we demonstrate the robustness of our procedure when applied to real data containing substantial amounts of outliers.

For comparison several indicators are introduced, two of which refer to a combination of classical procedures as benchmark.
To evaluate the performance of the $DD\alpha$-procedure under different depths and outsider treatments (when using the random Tukey depth) we have set up experiments with a large number of binary classification tasks with real data.
These data sets have been selected from open internet sources and different fields of application. Some of them have already served as benchmark sets in other classification studies. By this they are well suited for evaluating and comparing our new approach. The data can be downloaded in standardized form from
{\tt www.wisostat.uni-koeln.de/28969.html}. The complete $DD\alpha$-procedure is available as an R-package named {\tt ddalpha}.

The $DD\alpha$-classifiers are also compared with three traditional procedures: \emph{linear (LDA)} and \emph{quadratic (QDA) discriminant analysis} and the \emph{$k$-nearest-neighbors (KNN)} classifier, as in many cases (including the data sets considered here) at least one of them performs satisfactorily. We exclude neural network methods because they offer too many possible architectures, among which it is difficult to select in an automatic and computationally feasible way. While we expect that, given the specific data set, a properly adapted neural network performs rather well, such an approach affords a by hand tuning for each data set. Therefore we do not regard neural networks as fair competitors to the $DD\alpha$-classifiers.
We exclude as well the usual support vector machine (SVM) as a classifier because for each data set it has to be specially tuned.

Overview: Sect.~\ref{sec:ddaclassifier} describes the training phase of the $DD\alpha$-classifier, which consists of the depth transformation and the $\alpha$-separation in the extended depth space. The problem of generating directions for the random Tukey depth is discussed. Sect.~\ref{sec:treatments} regards the classification phase, where the problem of outsiders arises. Several classical approaches to classify the outsiders (LDA, maximum Mahalanobis depth, KNN) are introduced, as well as a simplified SVM approach, which liaises with the $DD\alpha$-separation in two ways.
Sect.~\ref{sec:realData} first describes the 50 classification tasks, which vary by absolute and relative sizes of training classes and include different portions of outliers and ties. Then the settings and results of the empirical study are presented.
In Sect.~\ref{sec:evidence} further evidence on outsider treatments and the number of (extended) depth properties needed is discussed. Sect.~\ref{sec:conclusions} concludes.

\section{Constructing the $DD\alpha$-classifier}\label{sec:ddaclassifier}

Consider a $q$-class classification problem, $q\ge 2$.
The $DD\alpha$-classifier has been recently proposed by  \cite{LangeMM12a}. Its classification phase consists of two parts: a transformation of the data from the original space into the {\em depth space} (depth transformation) and their subsequent separation using a modified version of the $\alpha$-procedure ($\alpha$-separation), see \cite{LangeM12}, \cite{Vasilev91}, \cite{Vasilev03}, \cite{VasilevL98}. This procedure is a projective invariant method to separate the depth transformed data.

\subsection{Depth transformation}\label{subsec:trans}

The \emph{depth transformation} maps $\mathbf{z} \in \mathbb{R}^d$ into $[0,1]^q$, the depth space, where the coordinates of the transformed data reflect their degree of centrality w.r.t. each of the $q$ classes, so that the subsequent class separation accounts only for the depth ordering. This central ordering is carried out using a properly chosen depth function $D(\cdot|\cdot)$. For more information on depth functions the reader is referred to the literature: e.g.\ \cite{ZuoS00} for properties and \cite{Mosler12} for a recent survey. Here we only briefly recall definitions of those used in the current work. The depth representation of the training sets in $[0,1]^q$ is called the \emph{depth plot}.

First we briefly regard three depths whose empirical versions take positive values beyond the convex hull of the data. For a point $\mathbf{z}\in \mathbb{R}^d$ and a random vector $X$ in $\mathbb{R}^d$ (especially one having an empirical distribution on a set of $d$-variate observations $\{\mathbf{x}_{1}, \dots ,\mathbf{x}_{n}\}$) the \emph{Mahalanobis depth} \citep{Mahalanobis36} of $\mathbf{z}$ w.r.t.\ $X$ is defined as
\begin{equation}\label{eqn:MahDepth}
D^{Mah}(\mathbf{z}|X) = (1 + (\mathbf{z} - \mu_{X})^{\prime}\Sigma_{X}^{-1}(\mathbf{z} - \mu_{X}))^{-1},
\end{equation}
where $\mu_{X}$ measures the location (e.g. the mean) of $X$, and $\Sigma_{X}$ the scatter (e.g. the covariance matrix) of $X$.

The \emph{affine invariant spatial depth} \citep{VardiZ00,Serfling02} of $\mathbf{z}$ regarding $X$ is defined as
\begin{equation}\label{eqn:sptDepth}
D^{Spt}(\mathbf{z}|X) = 1 - \| E_X \left[ v(\Sigma_{X}^{-1/2}(\mathbf{z} - X)) \right]\|\,.
\end{equation}
Here $v(\mathbf{y})=\|\mathbf{y}\|^{-1}\mathbf{y}$ for $\mathbf{y}\ne \mathbf{0}$ and $v(\mathbf{0})=\mathbf{0}$, and $\Sigma_{X}$ is the covariance matrix of $X$. As the Mahalanobis and spatial depths lack robustness when using standard moment estimates for $\Sigma_X$ and $\mu_X$, we consider also robustified versions of them, where $\Sigma_X$ and $\mu_X$ are estimated by MCD (minimum covariance determinant).

The \emph{projection depth} \citep{ZuoS00} of $\mathbf{z}$ regarding $X$ is given by
\begin{equation}\label{eqn:prjDepth1}
D^{Prj}(\mathbf{z}|X) = (1 + O^{Prj}(\mathbf{z}|X))^{-1},
\end{equation}
with
\begin{equation}\label{eqn:prjDepth2}
O^{Prj}(\mathbf{z}|X) = \sup_{\mathbf{u} \in S^{d-1}}\frac{\left|\mathbf{u}^{\prime}\mathbf{z} - m(\mathbf{u}^{\prime}X)\right|}{MAD(\mathbf{u}^{\prime}X)}\,,
\end{equation}
where $m(\mathbf{u}^{\prime}X)$ denotes the (univariate) median of $m(\mathbf{u}^{\prime}X)$ and $MAD(\mathbf{u}^{\prime}X)$ the (univariate) median of the absolute deviation of $\mathbf{u}^{\prime}X$ from its median.

The \emph{Tukey depth} or \emph{location depth} \citep{Tukey75,ZuoS00} of
$\mathbf{z}$ w.r.t.\ $X$ is defined as the minimal probability of $X$ lying in a halfspace bounded by a hyperplane through $\mathbf{z}$,
\begin{equation}\label{eqn:TukeyDepth}
D^{loc}(\mathbf{z}|X) = \inf{\{P(H):\,H\,\text{is a closed halfspace containing}\, \mathbf{z}\}},
\end{equation}
where $P$ is the probability distribution of $X$. (A closed halfspace is the set of all points that lie on one side of a hyperplane including that hyperplane.) If $P$ is an empirical distribution, this means that $D^{loc}(\mathbf{z}|X)$ is the minimal portion of the data that can be cut off by a hyperplane through $\mathbf{z}$.
Obviously, in general, the Tukey depth vanishes outside the convex hull of the distributions' support.

\begin{figure}[t!]
\begin{center}
	\includegraphics[keepaspectratio=true,scale=0.355]{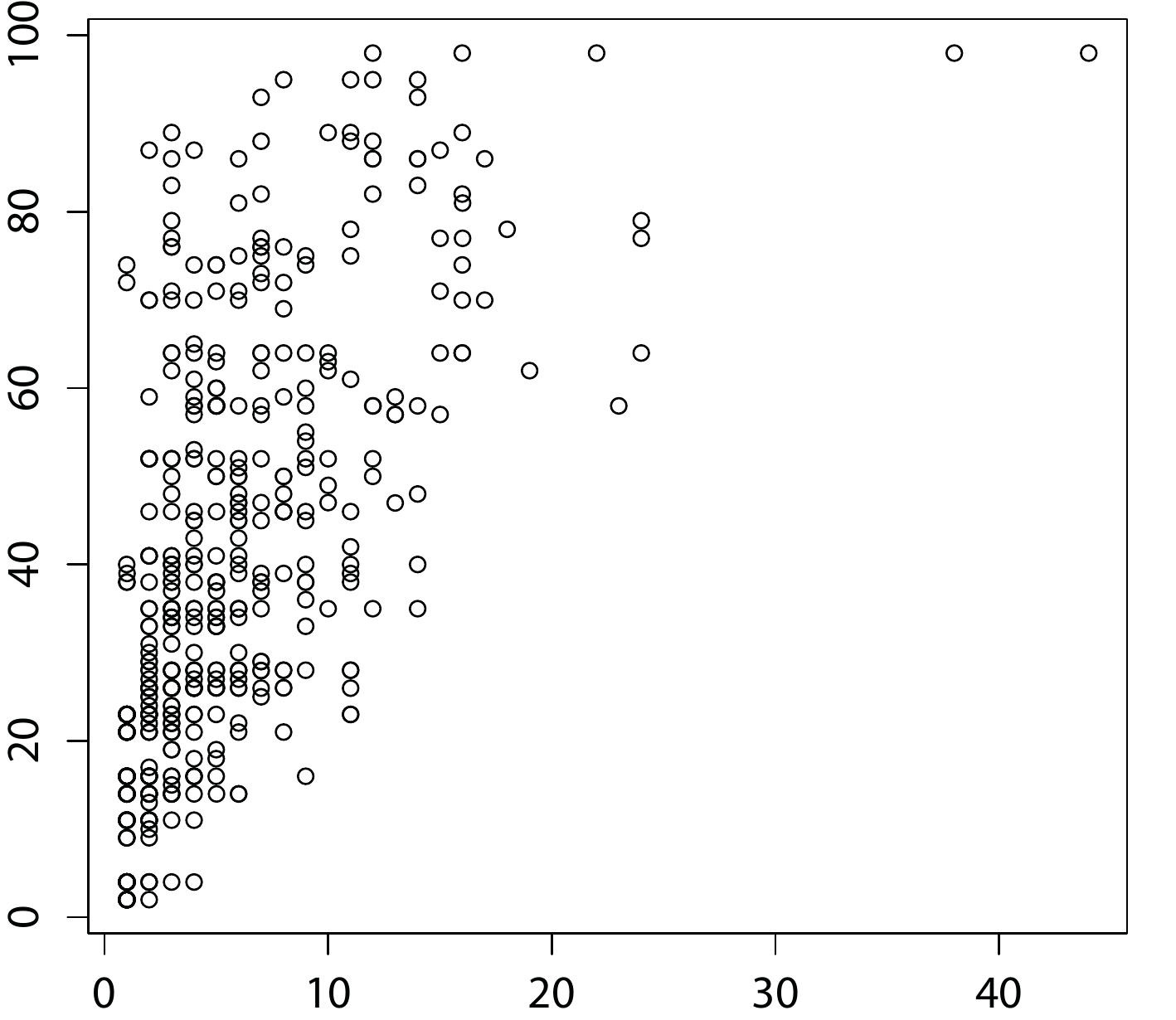}
	\includegraphics[keepaspectratio=true,scale=0.355]{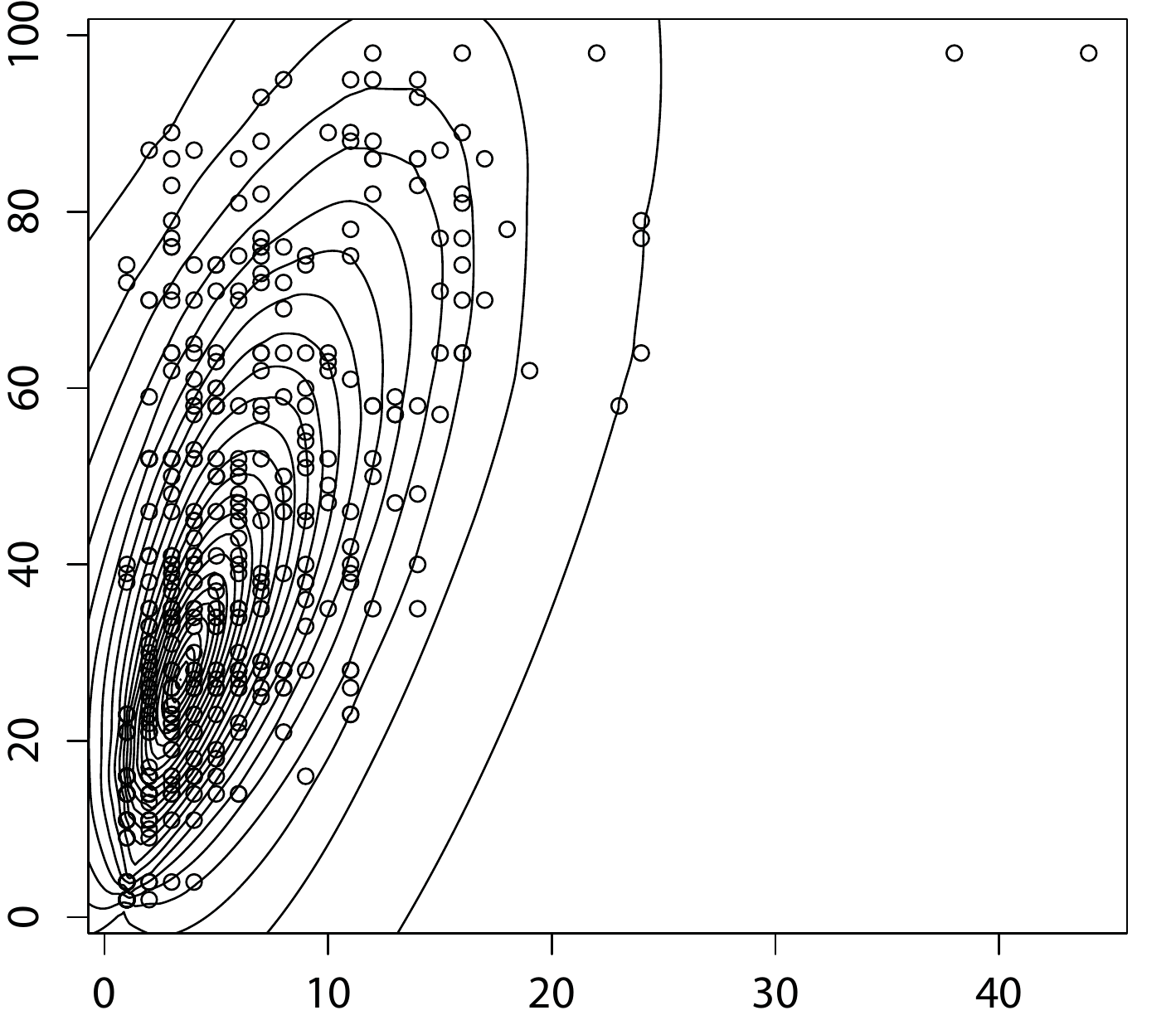}
	\includegraphics[keepaspectratio=true,scale=0.355]{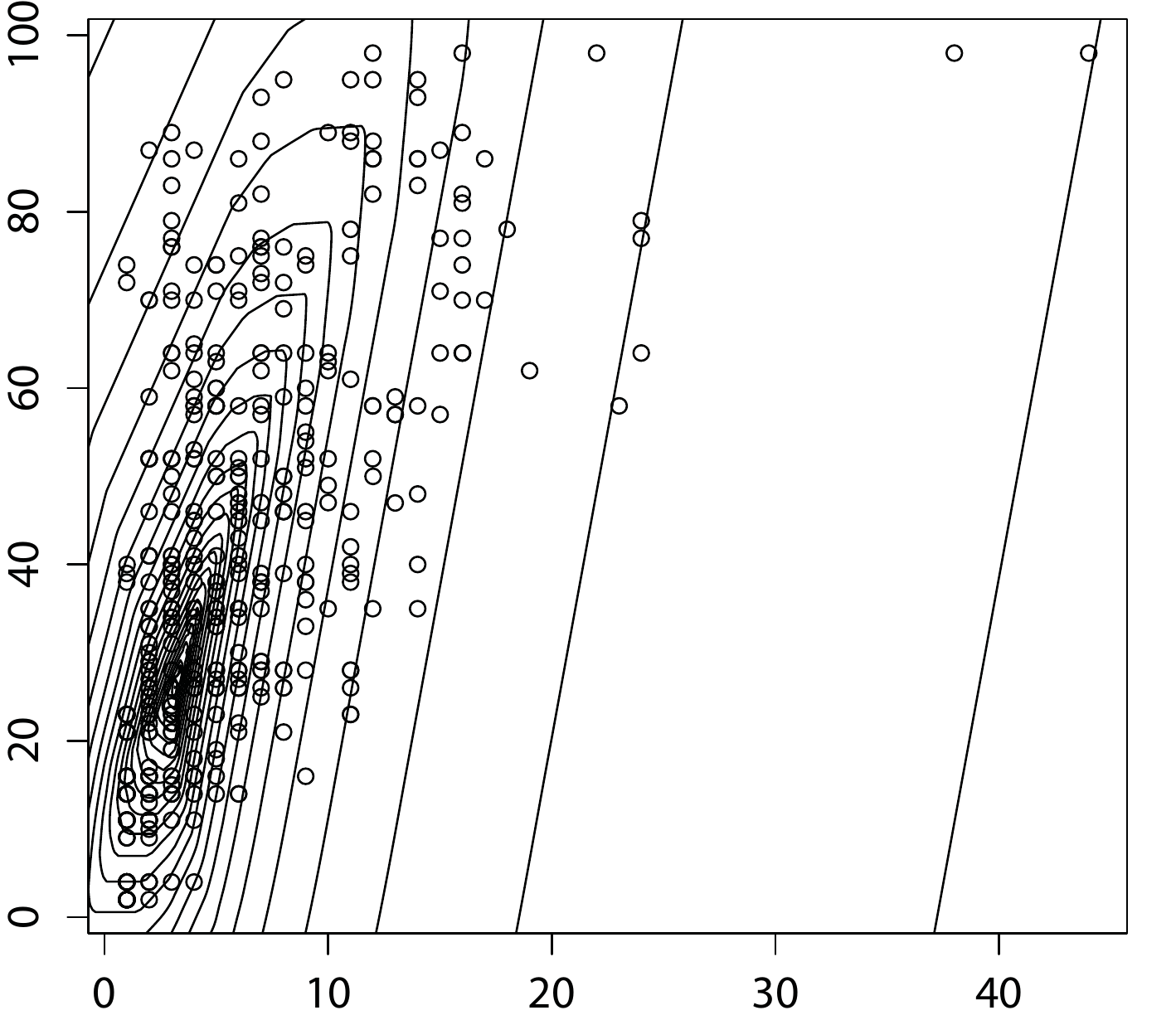}\\
	\includegraphics[keepaspectratio=true,scale=0.355]{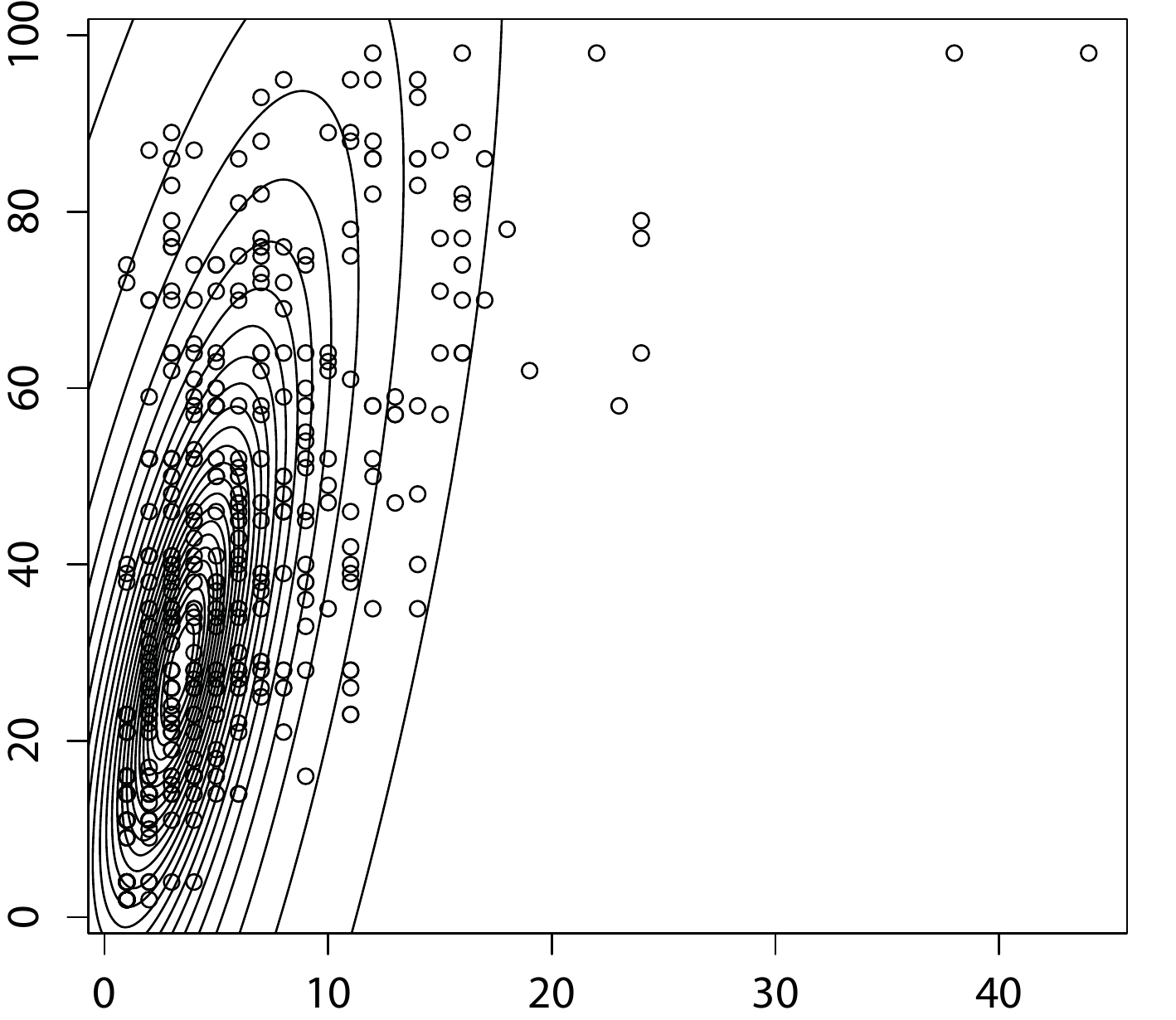}
	\includegraphics[keepaspectratio=true,scale=0.355]{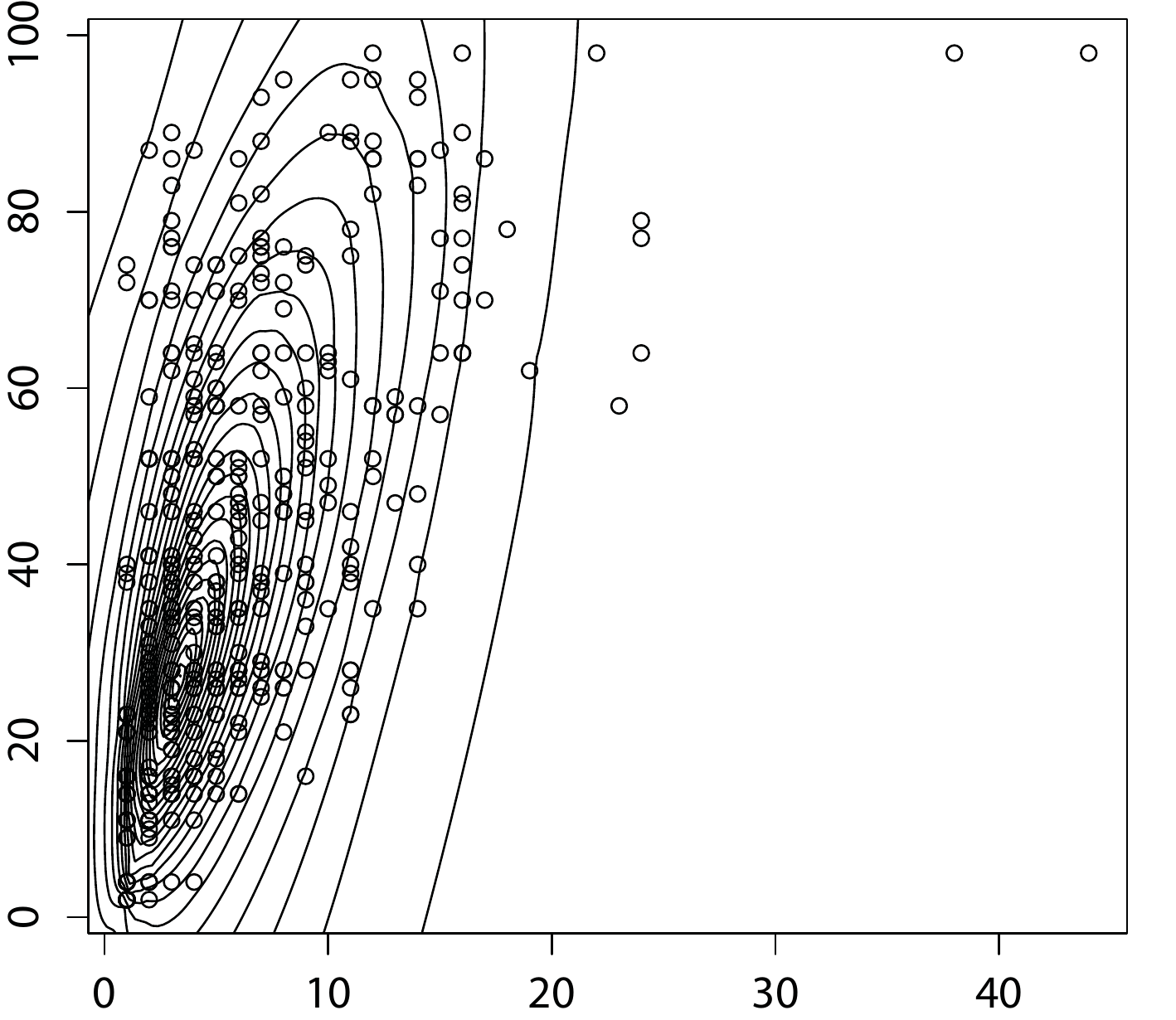}
	\includegraphics[keepaspectratio=true,scale=0.355]{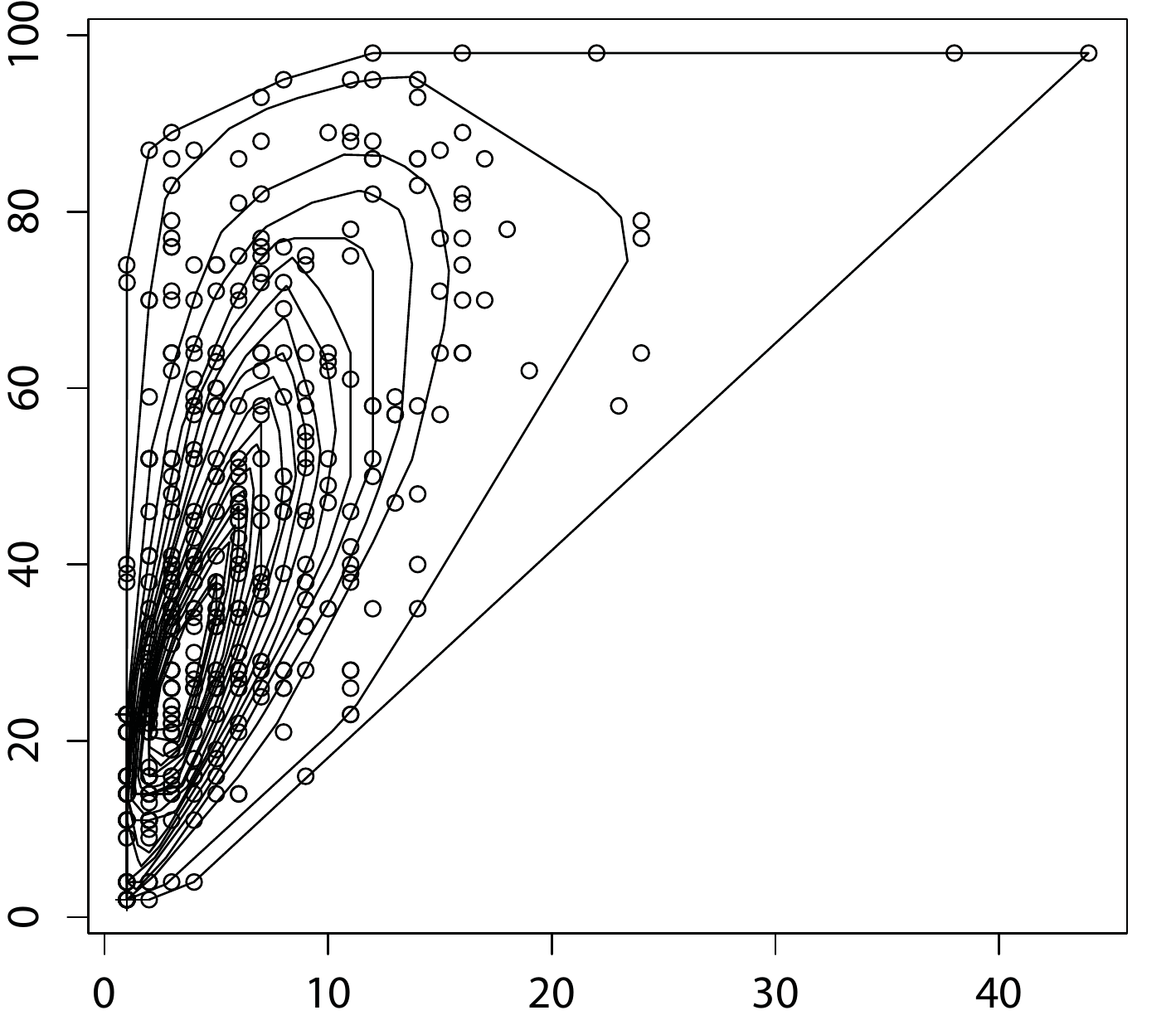}
	\caption{Depth level sets for two dimensions (total number of donations and months since first donation) of the ``blood-transfusion'' data set, class of not donating in March 2007. Top, from left to right: data; spatial depth using moment estimate; projection depth. Bottom: Mahalanobis depth using MCD estimates with robustness parameter $0.75$; spatial depth using the same MCD estimates; Tukey depth. The level sets are pictured for the depth values $1/570, 1/114, 1/57, 2/57, ..., 19/57$ for the Tukey depth, and for $0.04, 0.08, ..., 1$ for the rest of the depths.}
    \label{fig:depths}
\end{center}
\end{figure}

The Mahalanobis, spatial, projection depths are everywhere positive; thus outsiders cannot occur. However they are not very sensitive to the shape of the underlying distribution, which is illustrated in Figure~\ref{fig:depths}. It exhibits the data of the ``non-donating'' class in the ``blood-transfusion'' task.
The Figure shows the level sets of Mahalanobis, spatial, projection and Tukey depths.
Observe that the Mahalanobis depth yield ellipses, while with the spatial and projection depth rather symmetric, ellipsis-like level sets are obtained. The asymmetric shape of the data is much more reflected by the Tukey depth (bottom right); therefore it appears to be better suited to extracting the relevant information from the training classes.
Note that, by construction, the Mahalanobis depth has exact elliptical regions and the projection depth contains a symmetric factor, namely MAD, which accounts for the quasi-symmetric shape of the level sets. Moreover, the projection depth comes with an enormous computational cost.
For these reasons we include the Tukey depth in our study, in spite of its need for extra outsider treatments.


$D^{Mah}$ and $D^{Spt}$ are easily computed. To estimate $\mu_{X}$ (for $D^{Mah}$) and $\Sigma_{X}$ we use empirical moments and minimum covariance determinant (MCD) estimates that have outlyingness parameter $0.75$, see \cite{HubertD04}, \cite{RousseeuwD99}.
The Tukey depth as well as the projection depth satisfy the weak projection property \cite{Dyckerhoff04}, i.e. the depth of a point can be represented as the minimum of the depths on all unidimensional projections. Based on this we approximate the Tukey depth by the random Tukey depth \citep{CuestaANR08}, which is the minimum univariate Tukey depth  over a set of unidimensional projections in randomly selected directions.
As the exact calculation of $D^{Prj}$ is rather elaborate \citep{LiuZ12b}, we approximate it in the same way.

\subsection{$\alpha$-separation}\label{subsec:sep}

For each binary separation the \emph{$\alpha$-procedure} constructs a decision hyperplane in
the extended property space $E_1=[0,1]^r$, $r={p+q \choose q} - 1$. The extended property space includes the powers and products of objects' attributes up to some degree $p$ as additional coordinates. We mention the original depth coordinates as the \emph{basic}, and the other coordinates as the \emph{extended properties}. From all these properties, the relevant ones are stepwise chosen by the $\alpha$-procedure. The final separation then is performed by an `optimal' hyperplane in $[0,1]^r$.

Let $X_1$ and $X_2$ be two training classes in $\mathbb{R}^d$ having $n_1$ and $n_2$ elements respectively.
Each data point $\mathbf{x}\in X_1\cup X_2$ is transformed to $(d_1,d_2)\in [0,1]^2$, with $d_1=D(\mathbf{x}|X_1),d_2=D(\mathbf{x}|X_2)$.
In the {\em first step} all pairs of (basic or extended) properties are selected that involve depths in both classes and the respective two-dimensional coordinate subspaces are considered. In each of these planes we separate the two classes by a straight line passing through the origin. The separating line is determined by the angle $\alpha$ (formed with one of the axes) that yields the minimal \emph{empirical misclassification rate (EMR)}. Among all considered two-dimensional property spaces we choose the one delivering the smallest \emph{EMR}; then we project its points onto a straight line $f_1$ that is orthogonal to the separating one. Now, separation in $f_1$ is performed by the origin, and $f_1$ becomes the first \emph{projection axis} of a synthesized space.

We illustrate the procedure with data from the Pima Indians diabetes study; see {\tt www.stats.ox.ac.uk/pub/PRNN/pima.tr2}. A subsample consisting of $q=2$ classes (68 diseased and 132 not diseased females) has been selected, having seven attributes (number of pregnancies, 2 hours glucose concentration, blood pressure, triceps skin thickness, body mass index, diabetes pedigree function, age).
The data points are represented in the unit square by their random Tukey depths regarding the two classes (using 10~000 random directions), and powers and products of depth values are considered up to degree $p=2$; thus the extended depth space has dimension $r=5$.

\begin{figure}[t!]
\begin{center}
	\includegraphics{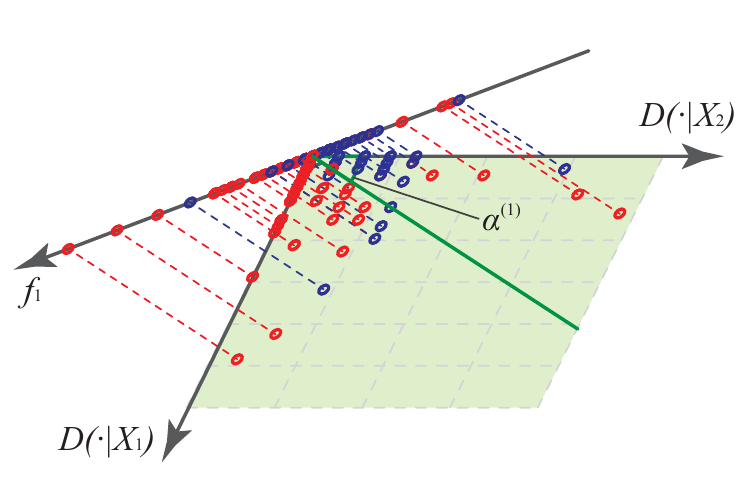}\,\,\,\includegraphics{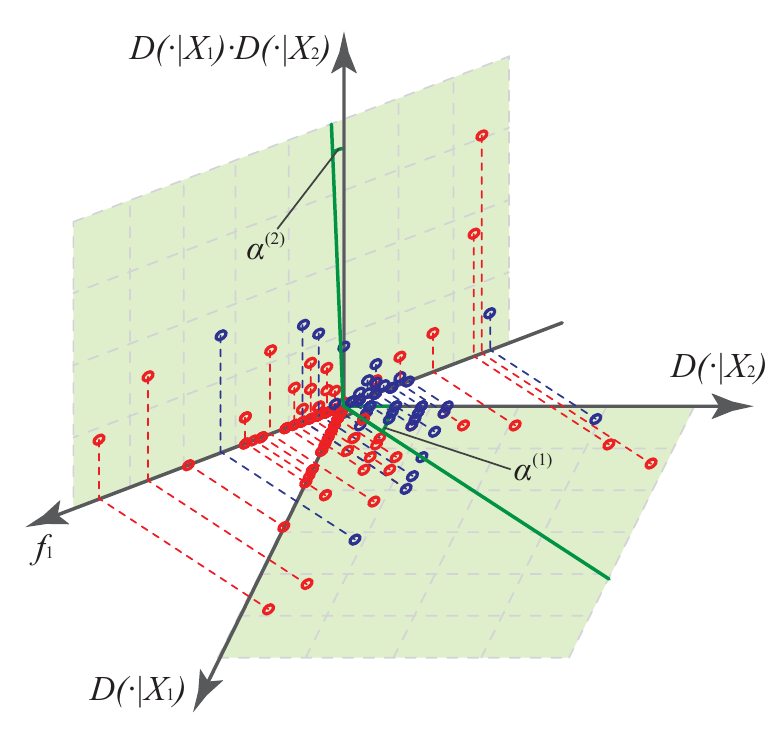}
	\caption{$\alpha$-separation; step 1 (left) and step 2 (right).}
    \label{figureAlpha}
\end{center}
\end{figure}

Figure \ref{figureAlpha} (left) shows the first step applied to the Pima data. Here, after mapping $X_1\cup X_2$ into $[0,1]^2$, the depth space was extended up to degree $p=2$, which yields the extended depth space $[0,1]^5$. With the Pima data the smallest \emph{EMR} is achieved at the two basic depth properties $d_1$ and $d_2$.

After the first step the extended space is reduced by removing the two (possibly extended) properties that have obtained minimal \emph{EMR}, and a similar second step is performed.
In the {\em second step} we consider all planar subspaces based on $f_1$ and one of the properties of $[0,1]^{r-2}$ and find a separating straight line minimizing \emph{EMR} in each of them. Again, among these planes we choose the one that yields the smallest \emph{EMR} and obtain the second projection axis $f_2$. It separates the data at its zero point, as the first axis did in the initial step; see Figure \ref{figureAlpha} (right) for the Pima example. Similarly, after Step 2 the extended space is reduced to an $(r-3)$-dimensional unit hypercube. The steps are iterated, and properties are selected from the extended space, as long as the minimum \emph{EMR} decreases and the remaining extended space is non-void. Note that with the Pima data, the procedure stops after Step 2.


\subsection{Directions for the random Tukey depth}\label{subsec:Tukey_depth}

As mentioned above,
we approximate the $d$-variate Tukey depth by the minimum univariate depth of the (on lines in several directions) projected data. The random Tukey depth inherits the robustness from the Tukey depth.
When implementing the random Tukey depth we have to answer the following questions in a way that makes our classifier reasonable and computationally feasible.
\begin{enumerate}
    \item[(1)] How should the random directions be generated?
    \item[(2)] How many directions should we consider?
    \item[(3)] Shall we generate a new set of directions for each point $\mathbf{z}$?
\end{enumerate}
Ad (1): As in \cite{CuestaANR08} we generate directions that are uniformly and independently distributed on the unit sphere and independent of the data.
Alternatively we could proceed as in \cite{ChristmannFJ02,ChristmannR01}, i.e.\ search through the normals of randomly formed $(d-1)$-simplices.
However we find it more efficient to spend computational time on generating random directions and evaluating univariate depths rather than computing exact directions from the data; moreover, a moderate number of random directions proves to be enough.

Ad (2): We mention two qualitative arguments concerning this number. Firstly, a larger dimension of the space needs a larger number of directions, which is obvious from the geometry of $\mathbb{R}^d$. However, we are not able to indicate the precise dependency of this number on the dimension $d$. Secondly,
we point out that there exists a trade-off between the number of directions used for the random Tukey depth and the number of outsiders. A simulation study that illuminates this trade-off has been conducted in \cite{LangeMM12b}. If the data stems from a centrally symmetric distribution (e.g. Gaussian or another elliptical distribution) the depth can be rather precisely approximated with a small number of directions. But if the data exhibit asymmetries and possible outliers, as real-world data mostly do, we need a larger number of directions to adequately represent them in the depth space. The degree of asymmetry and fat-tailedness found in the data may guide us in choosing this number; see \cite{LangeMM12b}.
To be able to compare the procedure on different data sets we have chosen a fixed number of directions.
This number has been set to 10~000 as a practical compromise between accuracy of the depth calculation and computational load.

It is obvious (see e.g. \cite{LiuZ12a}) that the random Tukey depth of a point can widely deviate from the Tukey depth.
On the other hand, the $\alpha$-separation is rather robust, since the only invariant it uses is whether a point belongs to a class or not.
Therefore an upward bias at a few points does not much influence the final separation.
Further, a trade-off exists between the number of directions used for the random Tukey depth and the number of effective outsiders, which favors a moderate number of directions.


Ad (3): We use the same set of directions for each data point. Though the generation of a new set of directions for each point produces different depth values, there is no reason to expect them to be more precise.
By keeping the set of directions constant we increase the speed of calculations in the training phase. The computations are boosted by avoiding the generation of the directions and projection of the points onto them. Let $k$ be the number of random directions. When instantly generating the direction set for each point the complexity of the depth calculation amounts to $\text{O}(kd(n_1+n_2)^2)$. If the direction set is not constantly changed, first, the time for their generation is saved. Second, recall that in the training phase the depth of all points of the training sample w.r.t. each class has to be computed. Hence, univariate projections are ordered, and all the univariate depths on each projection can be determined in a single pass.
This yields complexity $\text{O}(k(d+\log{(n_1+n_2)})(n_1+n_2))$. Note that $d(n_1+n_2)>d+\log{(n_1+n_2)}$ holds. Therefore, for all $d$, $n_1$ and $n_2$ that are large enough to suppress eventual constants, the constant direction approach is substantially faster, see also \cite{LangeMM12b}.
It also enhances the classifier's stability in the classification phase, as the same directions are used to approximate the depth of a new point to be classified.

\subsection{Directions for the projection depth}\label{subsec:projection_depth}

Exact calculation of the projection depth appears to be a heavy computational task \citep{LiuZ12b}.
Therefore we approximate the projection depth in the same way as the Tukey  depth by minimizing the univariate depth of projections in randomly drawn directions.
Compared with the Tukey depth, many more directions are needed to produce a reasonable approximation of the projection depth:
When traversing the unit sphere the projection depth (which is a piecewise linear function) changes direction much more frequently because of the median and MAD estimates. To be able to compare the procedure on different data sets we have chosen two fixed numbers of directions.
These have been set to 10~000 and 100~000 as a practical compromise between accuracy of the depth calculation and computational load.

\section{Classifying outsiders}\label{sec:treatments}

The \emph{classification phase} proceeds as follows.
Consider a point $\mathbf{z}$ to be classified. First the depth transform of $\mathbf{z}$ is calculated as $(d_1,d_2)=(D(\mathbf{z}|X_1),D(\mathbf{z}|X_2))$.
If $d_1$ is zero but not $d_2$, the point $\mathbf{z}$ is assigned to class $X_2$, and viceversa.
If both $d_1$ and $d_2$ are non-zero the point is classified according to the separation rule determined by the $\alpha$-procedure. This is always the case when using the Mahalanobis, spatial or projection depths.
When employing the random Tukey depth some $\mathbf{z}$ may have $(d_1,d_2)=(0,0)$, then $\mathbf{z}$ is regarded as an \emph{outsider}.
An outsider, being represented by the origin of the depth plot, cannot be readily classified but needs some special treatment.
Specifically, a point ${\mathbf{z}}$ that, in the original data space, lies outside the convex hulls of the two training sets has zero Tukey depth in both classes and, thus, is an outsider.
If a point ${\mathbf{z}}$ is no outsider
it is mentioned as an \emph{insider}. Insiders are instantaneously classified by the $\alpha$-procedure.

As in \cite{LangeMM12a}, outsiders may be classified by determining their nearest neighbors. In doing so, Euclidean and Mahalanobis distances can be employed, the latter to account for scatter within the classes.
Alternatively outsiders can be classified according to their maximum Mahalanobis depth, which is always positive.
\cite{HobergM06} introduce a depth function, which is the maximum of the zonoid depth and a properly scaled Mahalanobis depth, and thus circumvent the outsider problem.
\cite{PaindaveineVB12} propose an approach that avoids the outsider problem as well.
In the classification phase, for each point $\mathbf{z}$ to be classified, (1) the sample is extended by reflecting the training classes symmetrically at center $\mathbf{z}$, (2) the depth of points in the extended sample is considered, and (3) a $k$-nearest-neighbor rule that uses depth in place of distance is applied for classifying $\mathbf{z}$. Here not only the classification phase is computationally hard (by instantaneous calculation of the depths of all data points), but also the training phase, where the classifier has to be validated in order to determine $k$. This requires onerous computations.

In the sequel we compare several alternative outsider treatments, which are classifiers applied to data in the original space.
The treatments include three well known classifiers: linear discriminant analysis, maximum Mahalanobis depth, and $k$-nearest neighbors as well as a new one, which we call \emph{SVM-simplified}.
The performance of the $DD\alpha$-classifier with Mahalanobis, spatial and projection depth is contrasted as well.
Note that all depths (Tukey, Mahalanobis, spatial and projection) are affine invariant as well as all treatments used (LDA, KNN with an affine invariant distance, Mahalanobis depth and the support vector machine). Therefore all considered $DD\alpha$-classifiers are affine invariant (under appropriate moment assumptions), if the exact versions of the depths are calculated. Since the random Tukey depth and the random projection depth converge to the exact versions, using them makes the $DD\alpha$-classifiers approximately affine invariant.

The random Tukey depth (RTD) used for the depth transform is very efficiently calculated, but yields outsiders. As it approximates the Tukey depth (TD) from above, some TD-outsiders will have non-zero RTD and, by this, be assigned to one of the classes. A smaller number of directions yields a worse approximation of the TD, but reduces the number of outsiders. The remaining RTD-outsiders still need a special treatment, though.
Thus, when using the RTD, we face a trade-off between the quality of depth approximation and the extent of outsider treatment needed.
As we will see below with real data, choosing a moderate number of random directions gives best results.

\subsection{Classical approaches as treatments}\label{subsec:classtreatments}

\emph{Linear discriminant analysis} (LDA), introduced in \cite{Fisher36}, separates the classes by a hyperplane in the original data space; see also \cite{HastieTF09}. The LDA classifier is particularly simple. It is optimal if the data follow a Gaussian or, more general, a unimodal elliptical distribution and the classes differ by location shifts only. However, in classifying real data it is often outperformed by other approaches.
In many applications, the real data cannot be assumed to be Gaussian and ask for procedures different from LDA.
However, after having classified the RTD-insiders, the remaining task of classifying the outsiders appears to be a much less exigent task and may be  successfully done by a simple procedure like LDA.

The \emph{maximum-Mahalanobis-depth} classifier is given by
\begin{equation}\label{eqn:MahDepthDiscrimination}
\text{class} = \argmax_{i} \pi_i D^{Mah}(\mathbf{z}|X_i)\,,
\end{equation}
where
$\pi_i$ is the prior probability for class $i$.
The priors are estimated by the training class portions.
Again, this classifier has optimality properties under ellipticity.
Applied to outsiders it is expected to perform satisfactorily.

The \emph{$k$-nearest-neighbors} classifier is still another option for treating outsiders. Its parameter $k$, the number of the nearest neighbors, has to be chosen by cross-validation. Often a relatively small $k$ is enough; see, e.g., \cite{LangeMM12a}, where already $k=1$ produces satisfying results.
To make the procedure affine invariant we use Mahalanobis distances (based on the pooled data set) for finding nearest neighbors.

\subsection{SVM-simplified as an outsider treatment}\label{subsec:svm_method}

As another way to handle the outsider problem we propose to supplement the $DD\alpha$-classifier by an additional SVM-rule, which is restricted to classifying the outsiders. It has a particularly simple structure.
Recall that the $DD\alpha$-procedure delivers a separator which is a hyperplane in the extended depth space.
This hyperplane induces a decision rule in the original data space.
Next, we remove all training points which are not correctly classified by this rule (so that \emph{EMR}$\ =0$) and subject the remaining points to an additional SVM classification step that involves determining a single kernel parameter but no box-constraint. This new approach is named \emph{SVM-simplified} (\emph{SVM-s}).
As the \emph{SVM-s} rule is defined on the whole $\mathbb{R}^d$, it is able to assign points which are outsiders in the $DD\alpha$-classification.

\begin{figure*}[h!]
\begin{center}
	\includegraphics[keepaspectratio=true,scale=0.575]{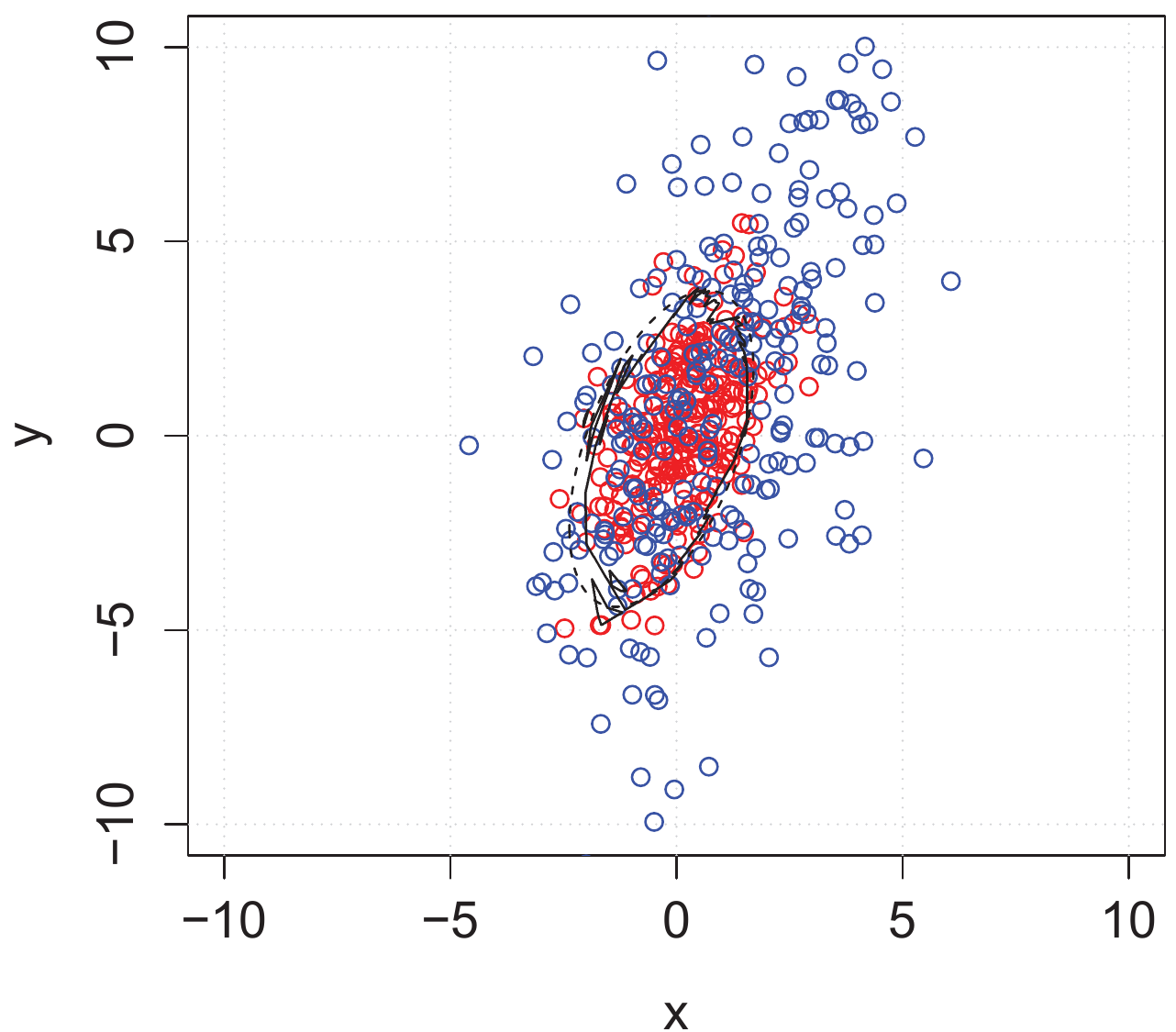}
    \includegraphics[keepaspectratio=true,scale=0.575]{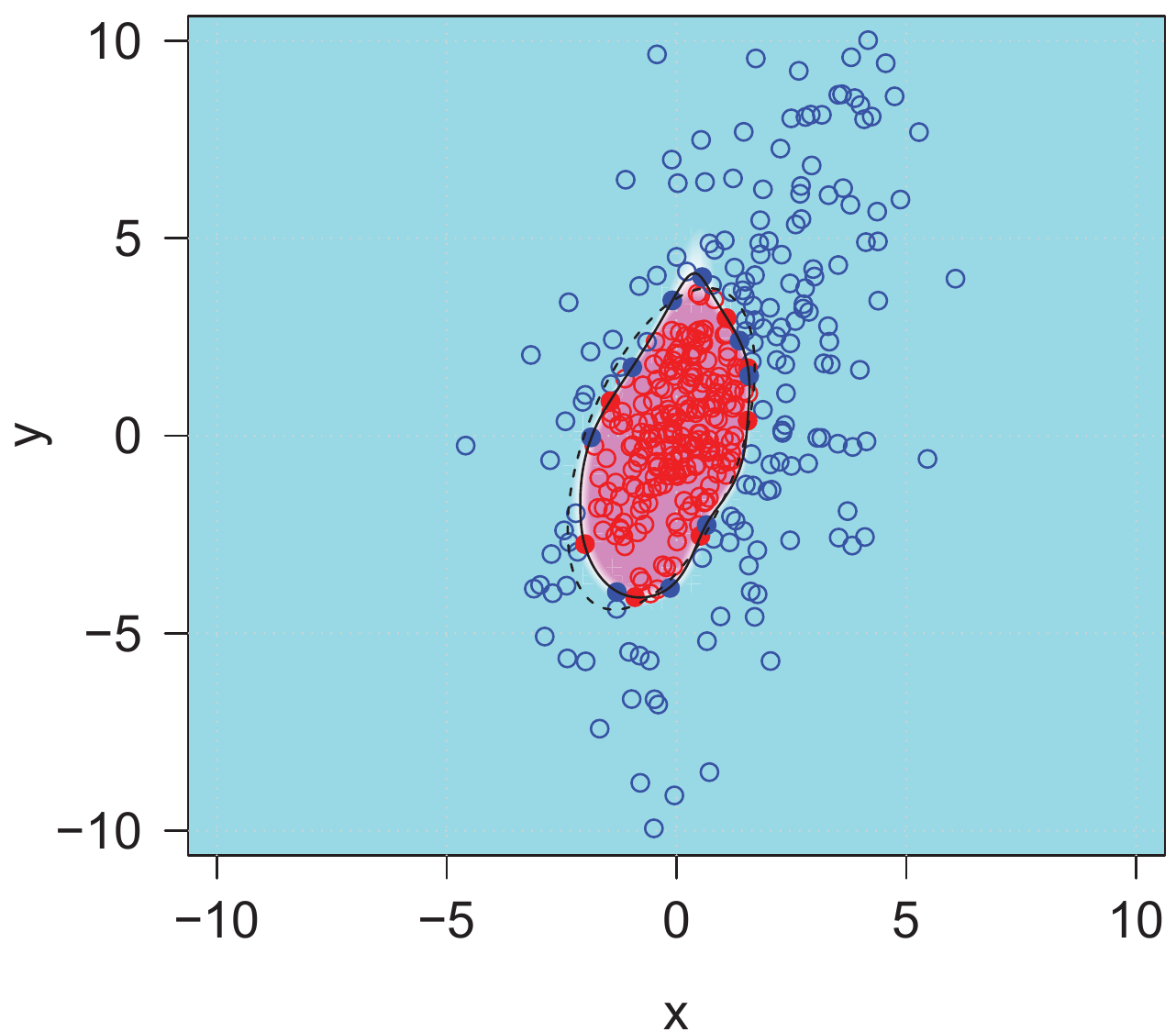}
	\caption{Decision rules of different classifiers: optimal Bayes (dashed lines), $DD\alpha$ (left, solid line), and SVM-simplified (right, solid line) classifiers.}
    \label{figSVM13}
\end{center}
\end{figure*}

Figure \ref{figSVM13}, left panel,  shows two classes, each containing 250 points, which are simulated from N$(\bigl[\begin{smallmatrix} 0\\0 \end{smallmatrix}\bigr],\bigl[\begin{smallmatrix} 1 & 1\\ 1 & 4 \end{smallmatrix}\bigr])$ (red) and N$(\bigl[\begin{smallmatrix} 1\\1 \end{smallmatrix}\bigr],\bigl[\begin{smallmatrix} 4 & 4\\ 4 & 16 \end{smallmatrix}\bigr])$ (blue), together
with the separating lines of the optimal Bayes (dashed) and $DD\alpha$ (solid) classifiers. The left panel regards the original data, while the right panel exhibits the data after the removal step.

The \emph{SVM-s} step consists in solving the following quadratic programming problem \citep{CortesV95}:
\begin{equation}\label{eqn:SVMproblem}
\maximize_{\lambda} \quad W(\mathbf{\lambda})=\mathbf{\lambda}^{\prime}\mathbf{1}-\frac{1}{2}\mathbf{\lambda}^{\prime}\mathbf{D}\mathbf{\lambda}
\end{equation}
subject to the constraints
\begin{equation}\label{eqn:SVMconditions1}
\mathbf{\lambda}\ge\mathbf{0},\\
\end{equation}
\begin{equation}
\mathbf{\lambda}^{\prime}\mathbf{y}=0.
\end{equation}
Here we notate $l=n_1+n_2$, $\mathbf{\lambda}=(\lambda_1,...,\lambda_l)'$,   $\mathbf{1}=(1,\dots,1)'$ and $\mathbf{0}=(0,\dots, 0)'\in \mathbb{R}^l$.
$\mathbf{y}$ stands for the $l$-dimensional vector of responses $y_1,...,y_l\in\{-1,1\}$,
and $\mathbf{D}$ is a symmetric $l\times l$-matrix with elements
\begin{equation}\label{eqn:SVMmatrix}
D_{ij}=y_iy_j K_\gamma(\mathbf{x}_i,\mathbf{x}_j),\text{   }i,j=1,...,l,
\end{equation}
where $K_{\gamma}(\mathbf{x}_i,\mathbf{x}_j)=\exp(-\gamma\|\mathbf{x}_i-\mathbf{x}_j\|^2)$ is a Gaussian kernel. Note that no box-constraint condition is needed here as the points are separable without error. But still the kernel parameter $\gamma$ has to be chosen.
For given $\gamma$ a solution $\mathbf{\lambda}^0=(\lambda_1^0,...,\lambda_l^0)$ of (\ref{eqn:SVMproblem}) is obtained, provided the two classes are linearly separable in the reproducing kernel Hilbert space that corresponds to $K_{\gamma}(\cdot,\cdot)$. Every such solution $\mathbf{\lambda}^0$ determines a margin between the classes $\rho_0=\sqrt{\frac{2}{W(\mathbf{\lambda}^0)}}$ and a number of support vectors, $\sharp\{\lambda_i^0|\lambda_i^0>0,i=1,...l\}$.

\begin{figure}[h!]
\begin{center}
	\includegraphics[keepaspectratio=true,scale=0.6]{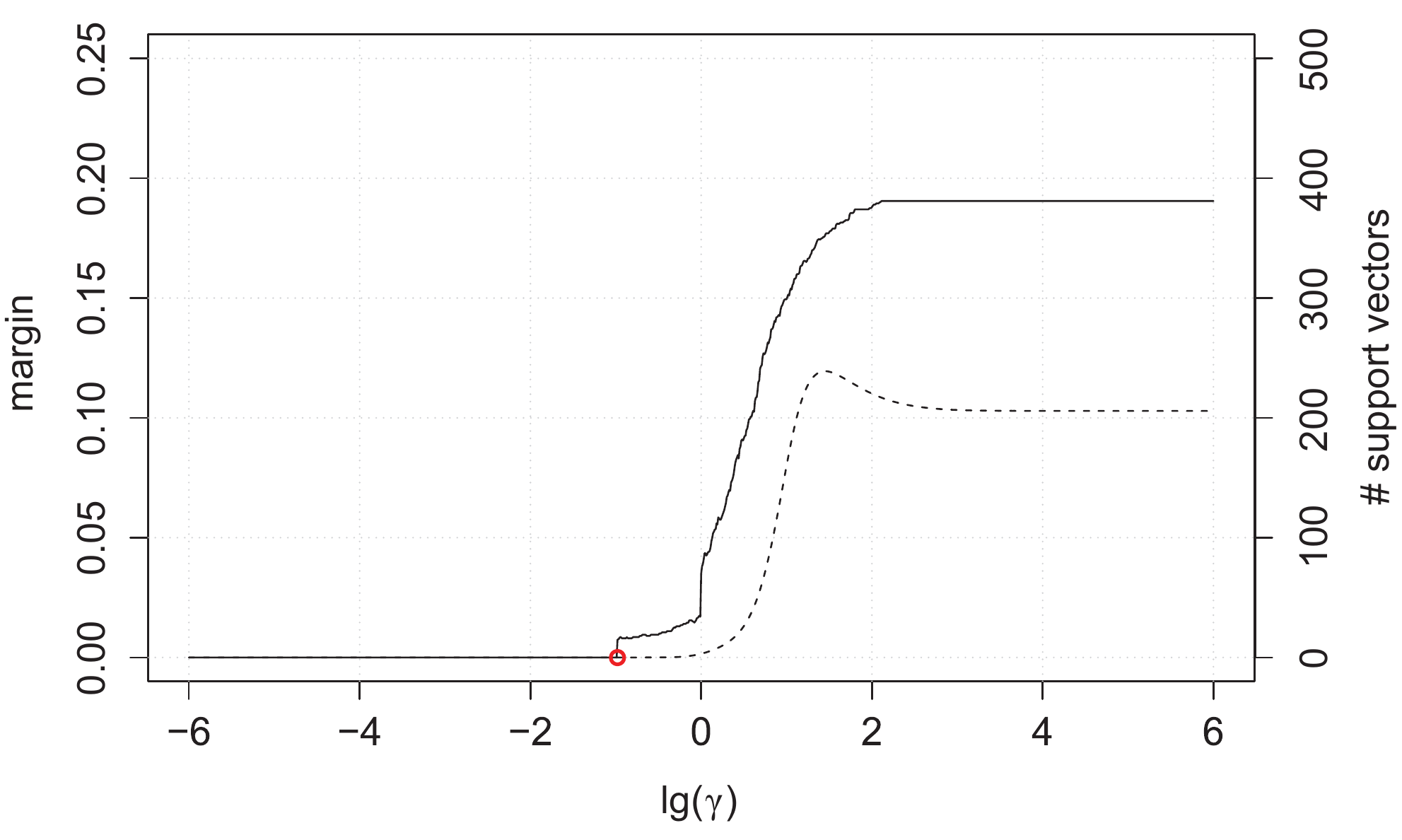}
	\caption{The number of support vectors (solid line, right scale) and the margin between the classes (dashed line, left scale) for different values of the parameter $\text{lg}(\gamma)$ (logarithm to the base 10).}
    \label{figSVM2}
\end{center}
\end{figure}

In Figure \ref{figSVM2}, depending on $\gamma$, the values of $\rho_0$ (dashed line) and the corresponding numbers of support vectors (solid line) are plotted for the above example. A zero value of the margin $\rho_0$ or of the number of support vectors
indicates that with the given $\gamma$ no errorless discrimination is possible.

Loosely speaking, $\gamma$ controls the complexity of the \emph{SVM-s} rule; for small values of $\gamma$ the decision rule is not able to separate the classes at all.
Therefore it suggests itself to use
the simplest separating rule, i.e.\ selecting the smallest $\gamma$ for which the classes are separated without error (as indicated by the small circle in Figure \ref{figSVM2}). This also comes out as a most stable decision rule. Figure \ref{figSVM13} exhibits the corresponding decision rule in its right panel as a solid line, while the dashed line indicates the same optimal Bayes decision rule as on the left panel of the Figure; the rules appear to be very similar.
Most important, calculating the \emph{SVM-s} rule needs no parameter tuning besides selecting the parameter $\gamma$, which is a straightforward task.

Needless to say, with the usual support vector machine (SVM) a solution can be obtained that is at least as good as the one achieved here.
However, obtaining this solution needs a tuning of parameters that is computationally much more intensive.

To summarize the above procedure: The training phase consists of two steps, first determining the $DD\alpha$-classifier and then determining an \emph{SVM-s} rule based on the correctly $DD\alpha$-classified points.
Note that the classification performance of this procedure is determined by the $DD\alpha$-classifier, and the \emph{SVM-s} step just extrapolates this classifier to treat the outsiders.
In our experiments the whole training phase took between a few seconds and several minutes of computation time (64-bit, 1 kernel of the iCore 7-2600 having enough operative memory). The time reached a maximum of 10 minutes with four very large data sets only.

In the classification phase we have two choices: Either using the obtained \emph{SVM-s} rule for all points ${\mathbf z}$ to be classified, or first check for each ${\mathbf z}$ whether it is an insider or an outsider and then classify it with the $DD\alpha$-rule if it is an insider and with the \emph{SVM-s} rule otherwise. The first choice yields a particularly fast procedure, as the \emph{SVM-s} rule does not involve any depth calculations in the classification phase. 
We choose the second one as it is in line with the application of the outsider treatments mentioned above. Its results are presented in column `SVM-s' of Tables \ref{tab:performance1} and \ref{tab:performance2}.

\section{Real data experiments}\label{sec:realData}

To evaluate the $DD\alpha$-procedures with different outsider treatments and to
judge their usefulness in practical applications
we have set up experiments based on a large variety of real data sets.
The methodology is applied to 50 binary classification tasks, which have been obtained from partitioning 33 freely accessible data sets, see Tables \ref{tab:datasets1} and \ref{tab:datasets2}.
The subset of the ``Pima Indian Diabetes'' described above is included in the Table as ``pima'' (No. 35).

The authors and introducers of the accessed data sets are
\cite{CoxJK82} (``biomed''),
\cite{MillerSVS79} (``cloud''),
\cite{GreanyK84} (``irish-ed''),
\cite{McGilchristA91} (``kidney''),
\cite{NierenbergSBDG89} (``plasma-retinol''),
\cite{BiblarzR93} (``socmob'') and
\cite{KalbfleischP80} (``veteran-lung-cancer''); these data sets have been downloaded from \\{\tt lib.stat.cmu.edu/datasets}.\\
Data sets ``chemdiab'' \citep{ReavenM79} and ``hemophilia'' \citep{HabemmaHVDB74}
have been taken from the {\tt R}-packages `locfit' and `rrcov' respectively.
The ``pima'' data set constitutes a training subsample of the ``diabetes'' (see below)
and can be downloaded from \\{\tt www.stats.ox.ac.uk/pub/PRNN} \citep{Ripley96}.\\
Datasets
``baby'',
``banknoten'' \citep{FluryR88},
``crab'' \citep{Ripley96},
``gemsen'' ,
``groessen'' \citep{Galton1885},
``tennis'',
``tips'' and
``uscrime'' \citep{HandDLMCO94}
have been downloaded from the teaching data base \\{\tt stat.ethz.ch/Teaching/Datasets}.\\
The rest of the data sets is taken from \\{\tt archive.ics.uci.edu/ml} \citep{FrankA10};\\
it in particular originates from \cite{YehYT09} (``blood-transfusion''),
\cite{WolbergM90} (``breast-cancer-wisconsin'') and
\cite{Turney93} (``vowel'').

Multiclass problems were reasonably split into binary classification problems, and some of the data sets were slightly processed by removing correlated attributes, by dropping objects with missing values, and by selecting prevailing classes.
For detailed descriptions of the data considered we refer to the corresponding literature and public repositories; the fifty tasks together with short descriptions of the data can be found on the web page \\{\tt www.wisostat.uni-koeln.de/28969.html}.


\subsection{Data}\label{subsec:datasetsDescription}


As we see from Tables \ref{tab:datasets1} and \ref{tab:datasets2}, the classification tasks are much different. The Tables show their basic parameters: dimension $d$ of the original space, log ratio of the cardinalities $n_1$ and $n_2$ of the training classes (so that the sign reflects which is larger), total sample length $n=n_1+n_2$, percentages of outliers and outsiders. As we see, up to 13 attribute dimensions are considered. The total sample sizes range from 47 to 1349, while the relative size of the two classes varies between 1 and 6.5.

\begin{sidewaystable}
\begin{center}
    \caption{Data set parameters; Part 1.}
    \label{tab:datasets1}
    \begin{tabular}{|r|l|r|r|r|r|r|r|r|}
        \hline
        No. & Data set & $n_1+n_2$ & $\ln(n_1/n_2)$ & $d$ & $(n_1+n_2)/d$ & \# tied & \% outl. & \% outs. \\ \hline\hline
          1 & baby & 247 & 0.63 & 5 & 49.4 & 0 & 5.3 & 31.6 \\ \hline
          2 & banknoten & 200 & 0.00 & 6 & 33.3 & 0 & 4.5 & 68.0 \\ \hline
          3 & biomed & 194 & -0.63 & 4 & 48.5 & 0 & 4.6 & 27.8 \\ \hline
          4 & blood-transfusion & 748 & -1.17 & 3 & 249.3 & 246 & 5.1 & 2.9 \\ \hline
          5 & breast-cancer-wisconsin & 699 & 0.64 & 9 & 77.7 & 236 & 8.3 & 46.9 \\ \hline
          6 & bupa & 345 & -0.33 & 6 & 57.5 & 4 & 8.4 & 44.3 \\ \hline
          7 & chemdiab\_1vs2 & 112 & -0.76 & 5 & 22.4 & 0 & 2.7 & 67.9 \\ \hline
          8 & chemdiab\_1vs3 & 69 & 0.09 & 5 & 13.8 & 0 & 2.9 & 81.2 \\ \hline
          9 & chemdiab\_2vs3 & 109 & 0.83 & 5 & 21.8 & 0 & 2.8 & 67.9 \\ \hline
         10 & cloud & 108 & 0.00 & 7 & 15.4 & 0 & 6.5 & 91.7 \\ \hline
         11 & crab\_BvsO & 200 & 0.00 & 5 & 40.0 & 0 & 2.0 & 57.5 \\ \hline
         12 & crab\_MvsF & 200 & 0.00 & 5 & 40.0 & 0 & 1.5 & 59.0 \\ \hline
         13 & crabB\_MvsF & 100 & 0.00 & 5 & 20.0 & 0 & 4.0 & 76.0 \\ \hline
         14 & crabF\_BvsO & 100 & 0.00 & 5 & 20.0 & 0 & 4.0 & 71.0 \\ \hline
         15 & crabM\_BvsO & 100 & 0.00 & 5 & 20.0 & 0 & 1.0 & 74.0 \\ \hline
         16 & crabO\_MvsF & 100 & 0.00 & 5 & 20.0 & 0 & 1.0 & 69.0 \\ \hline
         17 & cricket\_CvsP & 156 & 0.00 & 4 & 39.0 & 7 & 1.3 & 26.9 \\ \hline
         18 & diabetes & 768 & -0.62 & 8 & 96.0 & 0 & 8.9 & 56.8 \\ \hline
         19 & ecoli\_cpvsim & 220 & 0.62 & 5 & 44.0 & 0 & 5.9 & 42.3 \\ \hline
         20 & ecoli\_cpvspp & 195 & 1.01 & 5 & 39.0 & 0 & 5.1 & 43.1 \\ \hline
         21 & ecoli\_imvspp & 129 & 0.39 & 5 & 25.8 & 0 & 8.5 & 61.2 \\ \hline
         22 & gemsen\_MvsF & 1349 & 0.36 & 6 & 224.8 & 27 & 1.9 & 32.3 \\ \hline
         23 & glass & 146 & -0.08 & 9 & 16.2 & 1 & 11.6 & 91.1 \\ \hline
         24 & groessen\_MvsF & 230 & 0.02 & 3 & 76.7 & 0 & 2.6 & 20.4 \\ \hline
         25 & haberman & 306 & 1.02 & 3 & 102.0 & 23 & 2.9 & 7.5 \\ \hline
    \end{tabular}
\end{center}
\end{sidewaystable}
\begin{sidewaystable}
\begin{center}
    \caption{Data set parameters; Part 2.}
    \label{tab:datasets2}
    \begin{tabular}{|r|l|r|r|r|r|r|r|r|}
        \hline
        No. & Data set & $n_1+n_2$ & $\ln(n_1/n_2)$ & $d$ & $(n_1+n_2)/d$ & \# tied & \% outl. & \% outs. \\ \hline\hline
         26 & heart & 270 & -0.22 & 13 & 20.8 & 0 & 4.4 & 100.0 \\ \hline
         27 & hemophilia & 75 & -0.40 & 2 & 37.5 & 0 & 1.3 & 13.3 \\ \hline
         28 & indian-liver-patient\_1vs2 & 579 & 0.92 & 10 & 57.9 & 13 & 7.8 & 68.4 \\ \hline
         29 & indian-liver-patient\_MvsF & 579 & -1.14 & 9 & 64.3 & 13 & 7.9 & 54.7 \\ \hline
         30 & iris\_setosavsversicolor & 100 & 0.00 & 4 & 25.0 & 2 & 4.0 & 48.0 \\ \hline
         31 & iris\_setosavsvirginica & 100 & 0.00 & 4 & 25.0 & 3 & 4.0 & 48.0 \\ \hline
         32 & iris\_versicolorvsvirginica & 100 & 0.00 & 4 & 25.0 & 1 & 2.0 & 51.0 \\ \hline
         33 & irish-ed\_MvsF & 500 & 0.00 & 5 & 100.0 & 44 & 6.2 & 14.2 \\ \hline
         34 & kidney & 76 & -1.02 & 5 & 15.2 & 0 & 2.6 & 73.7 \\ \hline
         35 & pima & 200 & 0.66 & 7 & 28.6 & 0 & 5.0 & 86.0 \\ \hline
         36 & plasma-retinol\_MvsF & 315 & 1.87 & 13 & 24.2 & 0 & 8.3 & 98.1 \\ \hline
         37 & segmentation & 660 & 0.00 & 10 & 66.0 & 62 & 9.4 & 57.7 \\ \hline
         38 & socmob\_IvsNI & 1156 & 0.00 & 5 & 231.2 & 45 & 4.2 & 14.1 \\ \hline
         39 & socmob\_WvsB & 1156 & 0.00 & 5 & 231.2 & 8 & 3.0 & 15.1 \\ \hline
         40 & tae & 151 & -1.43 & 5 & 30.2 & 43 & 1.3 & 26.5 \\ \hline
         41 & tennis\_MvsF & 87 & -0.07 & 15 & 5.8 & 0 & 6.9 & 100.0 \\ \hline
         42 & tips\_DvsN & 244 & 0.95 & 6 & 40.7 & 1 & 5.3 & 48.8 \\ \hline
         43 & tips\_MvsF & 244 & -0.60 & 6 & 40.7 & 1 & 5.7 & 37.3 \\ \hline
         44 & uscrime\_SvsN & 47 & -0.65 & 13 & 3.6 & 0 & 0.0 & 100.0 \\ \hline
         45 & vertebral-column & 310 & 0.74 & 6 & 51.7 & 0 & 4.8 & 54.5 \\ \hline
         46 & veteran-lung-cancer & 137 & 0.01 & 7 & 19.6 & 0 & 8.8 & 80.3 \\ \hline
         47 & vowel\_MvsF & 990 & 0.13 & 13 & 76.2 & 0 & 2.1 & 99.7 \\ \hline
         48 & wine\_1vs2 & 130 & -0.19 & 13 & 10.0 & 0 & 3.8 & 100.0 \\ \hline
         49 & wine\_1vs3 & 107 & 0.21 & 13 & 8.2 & 0 & 0.9 & 100.0 \\ \hline
         50 & wine\_2vs3 & 119 & 0.39 & 13 & 9.2 & 0 & 3.4 & 100.0 \\ \hline
    \end{tabular}
\end{center}
\end{sidewaystable}

Almost all data contain \emph{outliers}; see column `\% outl.'.
In particular ``diabetes'', ``glass'', and ``segmentation'' contain substantial portions of them. The outliers of each data set have been identified by cutting moment Mahalanobis regions at a $\chi^2_d(0.975)$-quantile as, e.g., in \cite{RousseeuwD99}.
We take a pure data-analytic view and thus treat a potential outlier in the same way as any other point.
Observe that regarding eventual outliers the $DD\alpha$-procedure is highly robust for two reasons:
Firstly, the classification is done by the $\alpha$-procedure - a very robust approach - in a low-dimensional compact set, the unit cube of $\mathbb{R}^q$.
Secondly, a robust depth like the Tukey depth can be employed.

Nevertheless the Tukey-DD-classifier suffers from the existence of \emph{outsiders} as the Tukey depth vanishes outside the convex hulls of the training classes.
The performance of this procedure obviously depends on the portion of outsiders in the data.
We measure the \emph{outsider proneness} of a training set by the portion of points lying outside the convex hulls of all classes. I.e.\ for each point we check (leaving it out) whether it lies inside the convex hull of at least one class of the remaining training sample.
As shown in the two Tables (last column) the portion of outsiders varies from 0.029 to 1; see Sect.~\ref{subsec:disc_outsiders} for discussion.

An important parameter of a data set is the ratio of the sample size over the dimension, $(n_1+n_2)/d$. It relates
to the ability of the trained procedure to classify new data.
The ratio varies from $3.6$ 
to $249.3$. 

Finally, real data can contain ties, which require additional consideration by learning algorithms like KNN and depth based classifiers, and thus increase computation time. The number of tied points (in the pooled classes) is shown in column `\# tied' of both Tables. It is determined as the smallest number of points that has to be removed from the training sample to make the remaining ones pairwise distinct.

\subsection{Study settings}\label{subsec:studySettings}


In constructing the depth transform of the $DD\alpha$-classifier, the Mahalanobis depth (based on moment estimates as well as robust MCD estimates with outlyingness parameter 0.75) is computed exactly, while the projection depth is approximated using 10~000 and 100~000 random directions.
Instead of the exact Tukey depth we calculate the random Tukey depth (RTD) with 10~000 randomly chosen directions which are the same for all points of a given data set. Since the RTD approximates the Tukey depth from above and points having depth zero
in both classes are treated as outsiders, the number of outsiders is systematically underestimated when using the RTD.
These outsiders are treated with the techniques described in Sect.~\ref{subsec:classtreatments} and Sect.~\ref{subsec:svm_method}.

The $\alpha$-separation is performed in a polynomially extended depth space, where the degree of the polynomial is chosen by 50-fold cross-validation.
The complexity of the separation rule is, in a natural way, characterized by the dimension of the space needed.
See Sect.~\ref{subsec:disc_features} for results on the expected number of such features.

In classifying outsiders by KNN the number $k$ is selected by the same cross-validation strategy as with the traditional KNN classifier.
We determine the number $k$ of neighbors in the KNN classifier by leave-one-out cross-validation, performed over a wide range of neighborhoods, but - to save computation time - not over the whole sample size; the performance is still highly satisfactory.

The max-Mahalanobis-depth classifier is calculated either with moment or MCD estimates, setting $\alpha=0.75$.
As a basis for the SVM-simplified classifier Joachims's C++ implementation of SVM$^{light}$ \citep{Joachims99} is used with slight modifications and interfaced to the $R$-environment.

We use an $R$-implementation for the traditional KNN with ties broken at random; similarly when treating the outsiders by the affine invariant KNN.
In SVM-simplified ties are neglected.
The $\alpha$-procedure is tie-immune as well, but in contrast to SVM, it accounts for the number of tied points.




\subsection{Empirical comparison}\label{subsec:empComp}

\begin{figure*}[t!]
\begin{center}
	\includegraphics[keepaspectratio=true,scale=0.6]{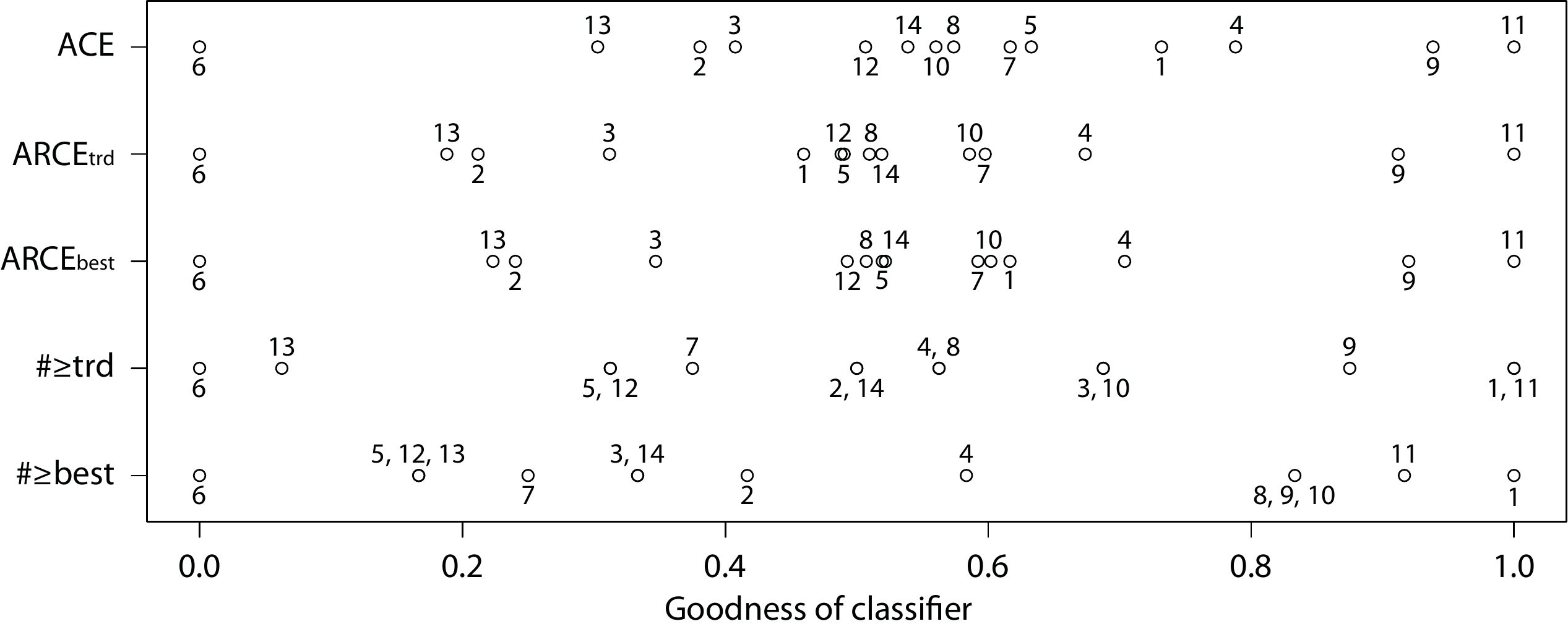}
	\caption{Goodness of the 14 considered classifiers (abscissa) w.r.t. the five performance measures (ordinate), obtained after standardizing the measures.
The classifiers are numerated as in Table~\ref{tab:indicators}.}
    \label{fig:results}
\end{center}
\end{figure*}

We solve the above fifty classification problems by the following fourteen approaches:
three classical approaches (LDA, QDA and KNN), the $DD\alpha$-classifier with the random Tukey depth and five outsider treatments from Sect.~\ref{sec:treatments} (LDA, KNN, maximum Mahalanobis depth classifier with both moment and robust estimates and SVM-s), the $DD\alpha$-classifier based on Mahalanobis, spatial (both using moment estimates and robust MCD-estimates with outlyingness parameter equal to 0.75), and projection depth (using 10~000 and 100~000 random directions).
The performance of each classifier is evaluated by leave-one-out cross-validation; we refer to this as the \emph{average error rate} (AER).
Tables \ref{tab:performance1} and \ref{tab:performance2} exhibit their  average error rates for each of the fifty settings, and for eleven classifiers. (For reasons of space, the $DD\alpha$-classifiers based on Mahalanobis and spatial depth with MCD-estimates and the one based on projection depth with 10~000 random directions are left out; short cumulative results are given in Table~\ref{tab:indicators} and Figure~\ref{fig:results}.)

The results are mixed. Some are surprising, e.g.\ the classification tasks ``bupa'', ``glass'', ``indian-liver-patient\_1vs2/FvsM'' show error rates for QDA that are substantially higher than those for LDA.
We attribute this to the poor estimation of the covariance matrix of the smaller class.

In Tables \ref{tab:performance1} and \ref{tab:performance2} classification error of the best classifier for each task is printed in bold.
In almost all tasks none of the considered procedures dominates the others. Exceptions are ``blood-transfusion'', ``indian-liver-patient\_FvsM'' and ``irish-ed\_MvsF'', where the
$DD\alpha$-classifier with the robust Mahalanobis depth (not shown in the Tables) dominates; also ``haberman'', where the $DD\alpha$-classifier with the projection depth using on 10~000 directions dominates, and ``cricket\_CvsP'',where they both are prevailing.


While at first sight the error rates of the diverse $DD\alpha$-classifiers show comparable sizes among each other and with the classical approaches (LDA, QDA, and KNN), it seems worthwhile to have a closer look at the performance of the different depths and treatments of outsiders.

Five aggregating measures are used to compare the overall performance of the considered classifiers.
The absolute performance is measured by the \emph{average classification error} (ACE), which is the average error of a classifier over all the classification tasks.
The relative measure is the \emph{average relative classification edge} (ARCE) calculated as the average of $((1-\text{AER})-(1-\text{AER}_\text{trd}))/(1-\text{AER}_\text{trd})$ over all classification tasks, with AER$_\text{trd}$ being the AER of the best of the three traditional classifiers (LDA, QDA and KNN) for each data set. We mention it as ARCE$_\text{trd}$. Its negative values relate to the best of the traditional classifiers, i.e. to an absolute reference. We also use ARCE with a relative reference (AER$_\text{best}$), which is the smallest AER among all considered classifiers for a given task (the bold values in Tables \ref{tab:performance1} and \ref{tab:performance2}); it depends on the variety of the classifiers chosen. The corresponding measure is mentioned as ARCE$_\text{best}$, which is always non-positive.
Two indicator measures denoted `\#$\ge \text{trd}$' and `\#$\ge \text{best}$' count how often AER of a classifier is not worse than AER$_\text{trd}$, respectively AER$_\text{best}$.

The five measures are given in Table~\ref{tab:indicators}, the best classifier w.r.t. each of the measures is printed in bold.
Note, that all proposed classifiers have negative ARCE$_\text{trd}$, i.e. none of them can outperform (on an average) the best of the traditional triplet.
On the other hand, the traditional classifiers perform mostly not satisfactory as well, although LDA shows favorable indicator values and competitive ACE.

To visualize the empirical evidence, the measures have been standardized to values in $\in [0,1]$, with larger numbers indicating better performance of the classifier, see Figure~\ref{fig:results}.
Three groups of the classifiers are easily distinguishable.
The first group consists mainly of the $DD\alpha$-classifier based on spatial depth followed by the one with Mahalanobis depth, both calculated using moment estimates. These two perform best. They also perform close to the best of the traditional classifiers (in terms of ARCE$_\text{trd}$, see Table~\ref{tab:indicators}), and not worse than this in (approximately) half of the cases (in terms of `\#$\ge \text{trd}$', see Table~\ref{tab:indicators}).
Results of the second group are mixed, only the $DD\alpha$-classifier based on the random Tukey depth supplemented with the LDA-treatment lies in parts on the positive border.

The $DD\alpha$-classifier based on random Tukey depth as transformation and moment-based Mahalanobis depth as outsider treatment performs worst. Similarly,
that based on projection depth with 10~000 random directions is mostly outperformed; this can be explained by insufficient approximation. (Note that with 100~000 random directions its performance increases).

\begin{sidewaystable}
\begin{center}
\caption{Comparison of classification performance (average error rate); Part 1.}
\label{tab:performance1}
\begin{tabular}{|r|l|r|r|r||r|r|r|r|r||r|r|r|}
\hline
    &         &     &     &     & \multicolumn{5}{|c||}{$DD\alpha$ + random Tukey depth + treatment} & \multicolumn{3}{|c|}{$DD\alpha$ + } \\
\cline{6-13}
    &         &     &     &     &     &     & \multicolumn{2}{|c|}{Mah.depth} & & \multicolumn{1}{|c|}{Mah.} & \multicolumn{1}{|c|}{Spt.} & \multicolumn{1}{|c|}{Prj.} \\
\cline{8-9}
No. & Dataset & LDA & QDA & KNN & LDA & KNN & Moment & MCD$_{0.75}$ & SVM-s & depth & depth & depth \\ \hline\hline
  1 & baby                      & 21.86 & 22.27 & 21.86 & \bf{21.46} & 25.10 & 25.51 & 23.48 & \bf{21.46} & 23.89 & 25.51 & 24.29 \\ \hline
  2 & banknoten                 & \bf{ 0.50} & \bf{ 0.50} & \bf{ 0.50} & \bf{ 0.50} &  1.00 &  4.00 & \bf{ 0.50} &  2.00 &  1.00 &  1.00 &  1.50 \\ \hline
  3 & biomed                    & 15.98 & 12.37 & 11.34 & 11.34 & 11.34 & 13.92 & \bf{10.82} & 11.86 & 12.37 & 13.40 & 13.40 \\ \hline
  4 & blood-transfusion         & 22.86 & 22.19 & 22.59 & 22.33 & 22.19 & 21.93 & 22.86 & 23.93 & 20.86 & 22.06 & 20.72 \\ \hline
  5 & breast-cancer-wisconsin   &  4.86 &  5.01 &  4.29 &  4.72 &  6.87 & 11.44 &  6.72 &  6.72 &  3.43 & \bf{ 3.00} &  4.43 \\ \hline
  6 & bupa                      & 30.72 & 40.58 & 31.30 & 26.96 & 28.70 & 30.43 & 31.88 & \bf{26.09} & 30.72 & 29.57 & 32.17 \\ \hline
  7 & chemdiab\_1vs2            &  3.57 &  7.14 &  7.14 &  3.57 &  5.36 & 14.29 &  5.36 & \bf{ 0.89} &  3.57 &  3.57 &  8.93 \\ \hline
  8 & chemdiab\_1vs3            & 10.14 &  8.70 &  7.25 &  8.70 &  8.70 & 17.39 & 13.04 & \bf{ 2.90} & 10.14 &  7.25 & 10.14 \\ \hline
  9 & chemdiab\_2vs3            &  3.67 &  0.92 &  0.92 &  3.67 &  6.42 &  1.83 &  0.92 & \bf{ 0.00} &  1.83 &  1.83 &  1.83 \\ \hline
 10 & cloud                     & 53.70 & 50.93 & 66.67 & 54.63 & 64.81 & \bf{40.74} & 48.15 & 59.26 & 51.85 & 49.07 & 49.07 \\ \hline
 11 & crab\_BvsO                & \bf{ 0.00} & \bf{ 0.00} &  3.00 & \bf{ 0.00} &  1.00 & \bf{ 0.00} & \bf{ 0.00} & \bf{ 0.00} & \bf{ 0.00} & \bf{ 0.00} & \bf{ 0.00} \\ \hline
 12 & crab\_MvsF                &  4.00 &  5.00 &  9.00 &  4.50 &  7.00 &  5.50 &  5.50 & \bf{ 3.50} &  4.50 & \bf{ 3.50} &  6.00 \\ \hline
 13 & crabB\_MvsF               &  9.00 & 10.00 & 15.00 &  8.00 &  9.00 & 10.00 &  8.00 &  7.00 & \bf{ 6.00} & \bf{ 6.00} &  9.00 \\ \hline
 14 & crabF\_BvsO               & \bf{ 0.00} &  1.00 &  5.00 & \bf{ 0.00} &  5.00 &  2.00 &  1.00 & \bf{ 0.00} &  1.00 &  1.00 &  2.00 \\ \hline
 15 & crabM\_BvsO               & \bf{ 0.00} & \bf{ 0.00} &  5.00 & \bf{ 0.00} &  1.00 & \bf{ 0.00} & \bf{ 0.00} & \bf{ 0.00} & \bf{ 0.00} & \bf{ 0.00} & \bf{ 0.00} \\ \hline
 16 & crabO\_MvsF               &  3.00 & \bf{ 2.00} &  7.00 &  3.00 &  6.00 &  3.00 &  3.00 &  3.00 & \bf{ 2.00} & \bf{ 2.00} & \bf{ 2.00} \\ \hline
 17 & cricket\_CvsP             & 68.59 & 64.10 & 59.62 & 64.74 & 67.31 & 64.10 & 62.82 & 57.05 & 61.54 & 57.69 & 58.33 \\ \hline
 18 & diabetes                  & \bf{22.53} & 26.04 & 24.61 & 26.04 & 25.52 & 29.04 & 27.21 & 27.34 & 23.57 & 24.22 & 34.24 \\ \hline
 19 & ecoli\_cpvsim             & \bf{ 1.36} &  1.82 &  2.27 &  1.82 &  2.73 &  5.91 &  1.82 &  4.55 & \bf{ 1.36} & \bf{ 1.36} &  2.73 \\ \hline
 20 & ecoli\_cpvspp             & \bf{ 3.08} &  4.10 &  4.10 &  4.62 &  4.62 &  5.64 &  6.15 &  9.23 &  4.62 &  5.13 &  5.13 \\ \hline
 21 & ecoli\_imvspp             &  5.43 &  3.88 &  5.43 &  5.43 &  4.65 &  9.30 &  6.20 &  5.43 & \bf{ 2.33} &  3.88 &  5.43 \\ \hline
 22 & gemsen\_MvsF              & 19.13 & 14.16 & 14.01 & 16.46 & 15.86 & 16.90 & 16.53 & 18.53 & 14.97 & \bf{13.94} & 23.28 \\ \hline
 23 & glass                     & 27.40 & 39.73 & \bf{19.18} & 29.45 & 27.40 & 34.93 & 30.82 & 28.77 & 30.14 & 28.08 & 39.73 \\ \hline
 24 & groessen\_MvsF            & 10.87 & 10.43 & 14.35 & 12.61 & 13.48 & 13.48 & 13.48 & 13.04 & 12.61 & \bf{ 7.83} & \bf{ 7.83} \\ \hline
 25 & haberman                  & 25.16 & 24.51 & 25.82 & 28.43 & 27.12 & 28.76 & 26.80 & 28.43 & 26.14 & 25.16 & 23.53 \\ \hline
\end{tabular}
\end{center}
\end{sidewaystable}
\begin{sidewaystable}
\begin{center}
\caption{Comparison of classification performance (average error rate); Part 2.}
\label{tab:performance2}
\begin{tabular}{|r|l|r|r|r||r|r|r|r|r||r|r|r|}
\hline
    &         &     &     &     & \multicolumn{5}{|c||}{$DD\alpha$ + random Tukey depth + treatment} & \multicolumn{3}{|c|}{$DD\alpha$ + } \\
\cline{6-13}
    &         &     &     &     &     &     & \multicolumn{2}{|c|}{Mah.depth} & & Mah. & Spt. & Prj. \\
\cline{8-9}
No. & Dataset & LDA & QDA & KNN & LDA & KNN & Moment & MCD$_{0.75}$ & SVM-s & depth & depth & depth \\ \hline\hline
 26 & heart                         & \bf{16.30} & 16.67 & 33.70 & 22.59 & 21.85 & 22.96 & 22.96 & 24.44 & 20.37 & 18.15 & 27.78 \\ \hline
 27 & hemophilia                    & \bf{14.67} & 16.00 & 16.00 & 16.00 & 17.33 & 18.67 & 18.67 & 18.67 & 17.33 & 16.00 & 21.33 \\ \hline
 28 & indian-liver-patient\_1vs2    & 29.88 & 44.56 & 31.26 & 30.92 & \bf{28.15} & 30.57 & 31.61 & 36.96 & 29.88 & 28.50 & 29.19 \\ \hline
 29 & indian-liver-patient\_FvsM    & 24.53 & 63.04 & 25.39 & 27.63 & 26.08 & 26.60 & 26.25 & 33.51 & 25.39 & 25.39 & 24.70 \\ \hline
 30 & iris\_setosavsversicolor      & \bf{ 0.00} & \bf{ 0.00} & \bf{ 0.00} & \bf{ 0.00} & \bf{ 0.00} & 12.00 & \bf{ 0.00} & \bf{ 0.00} & \bf{ 0.00} & \bf{ 0.00} & \bf{ 0.00} \\ \hline
 31 & iris\_setosavsvirginica       & \bf{ 0.00} & \bf{ 0.00} & \bf{ 0.00} & \bf{ 0.00} & \bf{ 0.00} & 12.00 & \bf{ 0.00} & \bf{ 0.00} & \bf{ 0.00} & \bf{ 0.00} & \bf{ 0.00} \\ \hline
 32 & iris\_versicolorvsvirginica   & \bf{ 3.00} &  4.00 & \bf{ 3.00} & \bf{ 3.00} & \bf{ 3.00} &  6.00 &  4.00 & 10.00 & \bf{ 3.00} &  6.00 &  5.00 \\ \hline
 33 & irish-ed\_MvsF                & 45.00 & 43.40 & 45.40 & 42.80 & 44.60 & 42.60 & 42.40 & 44.60 & 40.20 & 43.60 & 39.60 \\ \hline
 34 & kidney                        & \bf{28.95} & \bf{28.95} & 34.21 & \bf{28.95} & \bf{28.95} & 31.58 & 35.53 & 30.26 & \bf{28.95} & 31.58 & \bf{28.95} \\ \hline
 35 & pima                          & \bf{24.50} & 27.50 & 29.00 & 26.00 & 28.50 & 31.00 & 31.00 & 28.00 & 30.50 & 30.00 & 27.50 \\ \hline
 36 & plasma-retinol\_MvsF          & 14.29 & 13.97 & \bf{13.33} & 15.24 & 16.83 & 16.51 & 15.24 & 25.08 & 15.24 & 13.97 & 15.56 \\ \hline
 37 & segmentation                  &  8.33 &  9.24 &  4.55 &  4.55 & \bf{ 4.09} &  7.58 &  5.30 & 14.55 &  7.73 &  8.79 & 12.73 \\ \hline
 38 & socmob\_IvsNI                 & 34.34 & 34.34 & 33.48 & 32.70 & 33.30 & 32.61 & 33.13 & 33.04 & 32.09 & \bf{30.36} & 34.17 \\ \hline
 39 & socmob\_WvsB                  & 28.89 & 29.15 & 19.12 & 17.99 & 17.91 & \bf{17.65} & 18.08 & 18.94 & 20.16 & 19.64 & 21.89 \\ \hline
 40 & tae                           & 17.22 & 19.87 & 23.18 & \bf{11.92} & 13.91 & 14.57 & 17.22 & 15.89 & 17.22 & 16.56 & 17.22 \\ \hline
 41 & tennis\_MvsF                  & 41.38 & 44.83 & 43.68 & 45.98 & 49.43 & 47.13 & 39.08 & 52.87 & 37.93 & \bf{36.78} & 41.38 \\ \hline
 42 & tips\_DvsN                    &  6.15 &  3.69 &  8.20 &  5.33 &  6.15 & 12.70 &  9.84 &  8.20 & \bf{ 3.28} &  4.10 &  9.84 \\ \hline
 43 & tips\_MvsF                    & 36.48 & 38.52 & \bf{32.38} & 42.21 & 43.85 & 40.16 & 39.34 & 41.39 & 38.11 & 38.11 & 34.43 \\ \hline
 44 & uscrime\_SvsN                 & 17.02 & 19.15 &  8.51 & 17.02 &  6.38 & 48.94 & 21.28 & \bf{ 2.13} & 19.15 & 19.15 &  6.38 \\ \hline
 45 & vertebral-column              & 15.81 & 17.42 & 15.81 & 17.10 & 18.06 & 21.94 & 23.55 & 19.03 & \bf{14.52} & 15.16 & 16.45 \\ \hline
 46 & veteran-lung-cancer           & 64.23 & 51.82 & 51.82 & 53.28 & 43.80 & 48.91 & 42.34 & \bf{40.15} & 47.45 & 49.64 & 53.28 \\ \hline
 47 & vowel\_MvsF                   &  0.10 &  0.71 & \bf{ 0.00} &  1.41 &  1.92 & 11.82 &  0.00 &  2.02 &  0.51 &  0.51 & 12.63 \\ \hline
 48 & wine\_1vs2                    & \bf{ 0.00} &  0.77 &  6.15 &  1.54 &  2.31 &  5.38 &  2.31 &  3.08 &  1.54 &  1.54 &  1.54 \\ \hline
 49 & wine\_1vs3                    & \bf{ 0.00} & \bf{ 0.00} & 11.21 & \bf{ 0.00} &  0.93 &  0.93 & \bf{ 0.00} & \bf{ 0.00} & \bf{ 0.00} & \bf{ 0.00} & \bf{ 0.00} \\ \hline
 50 & wine\_2vs3                    &  0.84 & \bf{ 0.00} & 23.53 &  1.68 &  2.52 &  5.04 &  4.20 &  5.04 & \bf{ 0.00} & \bf{ 0.00} &  8.40 \\ \hline
\end{tabular}
\end{center}
\end{sidewaystable}
\begin{table}[t!]
\begin{center}
\caption{Comparison of classification indicators: average classification error (ACE); average relative classification edge (ARCE) w.r.t.\ the \emph{best traditional} and the \emph{overall best} classifier for a classification task; frequencies of better performance than the \emph{best traditional} (\#$\ge$trd) and the \emph{overall best} (\#$\ge$bst) classifier.}
\label{tab:indicators}
{\small
\begin{tabular}{|r|l||r|r|r|r|r|}
\hline
 N & Classifier & ACE & ARCE$_{\text{trd}}$ & ARCE$_{\text{bst}}$ & \#$\ge$trd & \#$\ge$bst \\ \hline\hline
 1 & LDA                                 & 16.79 & -2.95 & -4.73 & \bf{0.52} & \bf{0.32} \\ \hline
 2 & QDA                                 & 18.10 & -3.99 & -6.28 & 0.36 & 0.18 \\ \hline
 3 & KNN                                 & 18.00 & -3.57 & -5.84 & 0.42 & 0.16 \\ \hline
 4 & $DD\alpha$+RTD+LDA                  & 16.58 & -2.05 & -4.37 & 0.38 & 0.22 \\ \hline
 5 & $DD\alpha$+RTD+KNN                  & 17.16 & -2.82 & -5.13 & 0.30 & 0.12 \\ \hline
 6 & $DD\alpha$+RTD+Mah.d.(mom.)         & 19.52 & -4.88 & -7.27 & 0.20 & 0.08 \\ \hline
 7 & $DD\alpha$+RTD+Mah.d.(MCD$_{0.75}$) & 17.22 & -2.37 & -4.83 & 0.32 & 0.14 \\ \hline
 8 & $DD\alpha$+RTD+SVM-s                & 17.38 & -2.74 & -5.18 & 0.38 & 0.28 \\ \hline
 9 & $DD\alpha$+Mah.d.(mom.)             & 16.02 & -1.05 & -3.48 & 0.48 & 0.28 \\ \hline
10 & $DD\alpha$+Mah.d.(MCD$_{0.75}$)     & 17.43 & -2.42 & -4.79 & 0.42 & 0.28 \\ \hline
11 & $DD\alpha$+Spt.d.(mom.)             & \bf{15.79} & \bf{-0.68} & \bf{-3.15} & \bf{0.52} & 0.30 \\ \hline
12 & $DD\alpha$+Spt.d.(MCD$_{0.75}$)     & 17.63 & -2.83 & -5.24 & 0.30 & 0.12 \\ \hline
13 & $DD\alpha$+Prj.d.(10~000 dirs)      & 18.39 & -4.09 & -6.35 & 0.22 & 0.12 \\ \hline
14 & $DD\alpha$+Prj.d.(100~000 dirs)     & 17.51 & -2.70 & -5.12 & 0.36 & 0.16 \\ \hline
\end{tabular}
}
\end{center}
\end{table}

\section{Some evidence from the empirical study}\label{sec:evidence}

The empirical study sheds light on the nature and possible treatment of potential outsiders. It also provides practical evidence on the number of \emph{features} (= extended depth properties) that is normally needed in the linear separating procedure.

\subsection{Outsiders}\label{subsec:disc_outsiders}

It is seen from Tables \ref{tab:datasets1} and \ref{tab:datasets2} see that 10 (resp.\ 20\%) data sets contain more than 90\% outsiders, which actually means that less than 10\% of the points are classified by the $DD\alpha$-procedure, when the Tukey depth is employed. It is clear that this variant of the $DD\alpha$-procedure may be not the best solution, as it constructs a separation rule only from a small fraction of the training sample. In addition there are data sets (not that rarely encountered), where the outsider share amounts to 100\% so that the pure $DD\alpha$-approach does not separate anything. This also explains the result obtained in Section~\ref{subsec:empComp} that the better classifiers are based on non-vanishing depths.


\subsection{Dimension of the depth-feature space}\label{subsec:disc_features}

Finally, we come to the question, how many depth features are needed on an average to satisfactorily separate the data.
The $DD\alpha$-procedure uses the centered version of the $\alpha$-procedure \citep{Vasilev03,VasilevL98} to separate the classes in the depth space. As the $\alpha$-procedure is a heuristic approach, it is of interest how close this separation rule comes to the optimal one. Such a point of view corresponds to the theoretical assumption that the depth transformation does not impair the separability
of the data set. By its nature the resulting separating rule is similar to that proposed in \cite{LiCAL12}, where the polynomial degree is chosen by cross-validating. Other possible approaches are regression depth \citep{ChristmannR01,ChristmannFJ02} or SVM \citep{ChristmannFJ02,Vapnik98}. It is clear that in general the obtained separating hypersurface is not the one minimizing \emph{EMR},
 if more than two features are needed. But in which of the applications are they really needed?

Tables \ref{tab:features1} and \ref{tab:features2} exhibit the relative frequency of feature numbers, as they are
selected by $\alpha$-separation and  leave-one-out cross-validation. Surely these tables cannot be regard as histograms (since we would need some bootstrap for that), but it can be concluded that (on an average) in 99\% of the cases two features are sufficient for the separating rule. Note that the two-feature separators may include depth features of the extended type. Such a two-feature rule clearly minimizes the \emph{EMR} in the relevant plane.
Of course it may be possible to find a hyperplane in a more extended space, delivering a smaller \emph{EMR} than that obtained by the $\alpha$-separation. On the other hand, no space extension is needed in around 78\% of the cases. In these cases no polynomial products of depths are involved and the resulting separating rule is linear.

\renewcommand{\baselinestretch}{1.0}
\begin{table}[t!]
\begin{center}
\caption{Numbers of features selected by the $\alpha$-separation; Part 1.}
\label{tab:features1}
\begin{tabular}{|r|l|r|r|r|}
\hline
No. & Dataset & 2, \% & 3, \% & $\ge 4$, \% \\ \hline \hline
 1 & baby & 99.19 & 0.81 & 0.00 \\ \hline
 2 & banknoten & 100.00 & 0.00 & 0.00 \\ \hline
 3 & biomed & 100.00 & 0.00 & 0.00 \\ \hline
 4 & blood-transfusion & 91.31 & 1.34 & 7.35 \\ \hline
 5 & breast-cancer-wisconsin & 97.57 & 2.43 & 0.00 \\ \hline
 6 & bupa & 100.00 & 0.00 & 0.00 \\ \hline
 7 & chemdiab\_1vs2 & 100.00 & 0.00 & 0.00 \\ \hline
 8 & chemdiab\_1vs3 & 100.00 & 0.00 & 0.00 \\ \hline
 9 & chemdiab\_2vs3 & 100.00 & 0.00 & 0.00 \\ \hline
10 & cloud & 100.00 & 0.00 & 0.00 \\ \hline
11 & crab\_BvsO & 100.00 & 0.00 & 0.00 \\ \hline
12 & crab\_MvsF & 100.00 & 0.00 & 0.00 \\ \hline
13 & crabB\_MvsF & 100.00 & 0.00 & 0.00 \\ \hline
14 & crabF\_BvsO & 100.00 & 0.00 & 0.00 \\ \hline
15 & crabM\_BvsO & 100.00 & 0.00 & 0.00 \\ \hline
16 & crabO\_MvsF & 100.00 & 0.00 & 0.00 \\ \hline
17 & cricket\_CvsP & 98.72 & 1.28 & 0.00 \\ \hline
18 & diabetes & 99.61 & 0.13 & 0.26 \\ \hline
19 & ecoli\_cpvsim & 100.00 & 0.00 & 0.00 \\ \hline
20 & ecoli\_cpvspp & 100.00 & 0.00 & 0.00 \\ \hline
21 & ecoli\_imvspp & 100.00 & 0.00 & 0.00 \\ \hline
22 & gemsen\_MvsF & 99.85 & 0.07 & 0.07 \\ \hline
23 & glass & 100.00 & 0.00 & 0.00 \\ \hline
24 & groessen\_MvsF & 99.57 & 0.00 & 0.43 \\ \hline
25 & haberman & 100.00 & 0.00 & 0.00 \\ \hline
 & Average & 99.08 & 0.65 & 0.27 \\ \hline
\end{tabular}
\end{center}
\end{table}
\renewcommand{\baselinestretch}{1.5}

\renewcommand{\baselinestretch}{1.0}
\begin{table}[t!]
\begin{center}
\caption{Numbers of features selected by the $\alpha$-separation; Part 2.}
\label{tab:features2}
\begin{tabular}{|r|l|r|r|r|}
\hline
No. & Dataset & 2, \% & 3, \% & $\ge 4$, \% \\ \hline \hline
26 & heart & 100.00 & 0.00 & 0.00 \\ \hline
27 & hemophilia & 100.00 & 0.00 & 0.00 \\ \hline
28 & indian-liver-patient\_1vs2 & 99.48 & 0.52 & 0.00 \\ \hline
29 & indian-liver-patient\_MvsF & 98.96 & 1.04 & 0.00 \\ \hline
30 & iris\_setosavsversicolor & 100.00 & 0.00 & 0.00 \\ \hline
31 & iris\_setosavsvirginica & 100.00 & 0.00 & 0.00 \\ \hline
32 & iris\_versicolorvsvirginica & 100.00 & 0.00 & 0.00 \\ \hline
33 & irish-ed\_MvsF & 99.60 & 0.20 & 0.20 \\ \hline
34 & kidney & 98.68 & 1.32 & 0.00 \\ \hline
35 & pima & 93.00 & 5.50 & 1.50 \\ \hline
36 & plasma-retinol\_MvsF & 99.37 & 0.63 & 0.00 \\ \hline
37 & segmentation & 93.48 & 4.85 & 1.67 \\ \hline
38 & socmob\_IvsNI & 99.57 & 0.35 & 0.09 \\ \hline
39 & socmob\_WvsB & 99.83 & 0.09 & 0.09 \\ \hline
40 & tae & 93.38 & 6.62 & 0.00 \\ \hline
41 & tennis\_MvsF & 100.00 & 0.00 & 0.00 \\ \hline
42 & tips\_DvsN & 99.59 & 0.41 & 0.00 \\ \hline
43 & tips\_MvsF & 100.00 & 0.00 & 0.00 \\ \hline
44 & uscrime\_SvsN & 100.00 & 0.00 & 0.00 \\ \hline
45 & vertebral-column & 98.39 & 1.61 & 0.00 \\ \hline
46 & veteran-lung-cancer & 100.00 & 0.00 & 0.00 \\ \hline
47 & vowel\_MvsF & 94.75 & 3.43 & 1.82 \\ \hline
48 & wine\_1vs2 & 100.00 & 0.00 & 0.00 \\ \hline
49 & wine\_1vs3 & 100.00 & 0.00 & 0.00 \\ \hline
50 & wine\_2vs3 & 100.00 & 0.00 & 0.00 \\ \hline
 & Average & 99.08 & 0.65 & 0.27 \\ \hline
\end{tabular}
\end{center}
\end{table}
\renewcommand{\baselinestretch}{1.0}

\section{Conclusions}\label{sec:conclusions}

A fast classification procedure, the $DD\alpha$-procedure, has been introduced that is essentially nonparametric, robust, and computationally feasible for any dimension $d$ of attributes. The $DD\alpha$-procedure is available in the R-package {\tt ddalpha}.
The $DD\alpha$-classifier is particularly robust for two reasons:
first, as the classification is done by the $\alpha$-procedure, which is \emph{per se} robust;
second, as the data are transformed into a low-dimensional compact space.
Generally, two cases are to be distinguished with the depth transform: non-vanishing depth or depth vanishing beyond the convex hulls of the training classes.
Non-vanishing depths, e.g.\ Mahalanobis or spatial, often induce a spurious symmetry and are intrinsically non-robust (though, can be robustified). The projection depth, which is non-vanishing, also produces ellipsis-like regions. It is robust, but computationally inefficient when $d\ge3$. The last problem is faced by the Tukey depth, which best reflects the shape of the data. In place of the exact versions of projection and Tukey depth we employ their random versions by minimizing univariate depths in directions that are uniformly distributed on the sphere. A very large number of these directions is needed for the calculation of the projection depth, while for the random Tukey depth the number of directions can be kept low, as there is a tradeoff between this number and the number of points being classified by their depth values.

On the other hand the random Tukey depth yields outsiders when classifying, i.e. points lying outside the convex hulls of all classes, which cannot be readily classified and need additional treatment. In real data applications the percentage of outsiders can be large (see the introduced measure of outsider proneness) and thus substantially influence the classification performance. Therefore, the choice of the treatment is important when applying the random Tukey depth. The treatments considered subject the outsiders either to linear discriminant analysis (LDA), classification according to $k$-nearest neighbors (KNN), maximum Mahalanobis depth classification based on moment or MCD estimates, or the newly introduced SVM-simplified procedure (\emph{SVM-s}). The latter is very fast as it needs no tuning of a box-constraint; only the smallest separating $\gamma$ has to be computed. Additional calculations (not included here, see also \cite{LangeMM12a}) show, that regarding the other possible outsider treatments, the choices of number $k$ in KNN as well as of the covering parameter in MCD do not much influence their performance.

Thus the $DD\alpha$-procedure needs practically no parameter tuning.
The degree of the separating polynomial is chosen by cross-validation within the depth representation only, where the modified $\alpha$-procedure, on each of the planar subspaces considered, has a quick sort complexity, O$((n_1+n_2)\log (n_1+n_2)$, and by that is very fast.

The above introduced variants of the $DD\alpha$-procedure are challenged by 50 binary classification problems that arise from a broad range of real data. The tasks are complicated by the presence of outliers and ties.
As competitors of the $DD\alpha$-procedure  three traditional classification methods (LDA, QDA, and KNN) are evaluated with the same data.
Naturally, none of the classifiers is best at all tasks in terms of the average error rate, but each classifier is best at some of them.
Our results also show that no single depth or outsider treatment dominates the others.
Just for almost all data sets the classification of outsiders according to their maximum moment-Mahalanobis depth is outperformed by the same with the MCD-Mahalanobis depth. This can be explained by the outliers present.

Five goodness measures are introduced aggregating performance of the classifiers over the 50 classification tasks w.r.t. different aspects and allowing for direct comparison of the classifiers. Clearly, classification performance greatly depends on the choice of the depth and, if needed, the outsider treatment. The measures point out two $DD\alpha$-classifiers (starting with the best): based on spatial and Mahalanobis depth (both using moment estimates). The rest of the classifiers show varying performance for different goodness measures, which is demonstrated by the goodness visualization.

The experience with real-data problems tells us further that, in most practical cases, the separation procedure in the depth space stops after a few steps. In most cases the subspace spanned by the depth features needed has dimension two, which points out the high stability of the separating rule.

The problems and solutions investigated in this study are also of interest in more general settings:
The problem of coping with outsiders is common to any statistical procedure that is based on depth plots and involves a notion of depth vanishing outside the convex hull.
Using the random Tukey depth as an efficient approximation of the Tukey depth and selecting the random directions has many applications.
Available algorithms for exactly calculating the Tukey depth \citep{RousseeuwS98,LiuZ12a} are computationally expensive, but can serve as a benchmark.
Finally, the SVM-simplified method is introduced and appears as a simple and efficient way to avoid the computational burden of tuning the SVM.

\section*{Acknowledgements}

We thank the maintainers and contributors of the repositories \\{\tt archive.ics.uci.edu/ml},\\
{\tt lib.stat.cmu.edu/datasets},\\
{\tt stat.ethz.ch/Teaching/Datasets}.

We are also grateful to Oleksii Pokotylo, master student at the National Technical University of Ukraine, for his help in preparation and maintaining the R-package {\tt ddalpha}.
Valuable comments of two anonymous referees and an editor are greatly acknowledged.

\end{document}